\RequirePackage{fix-cm}
\documentclass[smallextended]{svjour3}                     
\usepackage{bm,subfigure,comment,amssymb,amsmath}
\usepackage[english]{babel}
\usepackage{color}
\usepackage{slashed}

\usepackage{wasysym}
\usepackage{latexsym}
\usepackage{graphicx}
\usepackage{fancyhdr}
\usepackage{color}
\usepackage{txfonts}

\usepackage{a4wide}
\newcommand{\beq}{\begin{equation}}
\newcommand{\eeq}{\end{equation}}
\definecolor{darkgray}{gray}{.6}
\def\makeheadbox{}

\begin{document}

\title{Efimov physics from the functional renormalization group}


\author{Stefan Floerchinger\and Sergej Moroz         \and Richard Schmidt
}


\institute{S. Floerchinger \at
              Institut f\"ur Theoretische Physik, Philosophenweg 16, D-69120 Heidelberg, Germany
              \\
             \emph{now at:} Physics Department, Theory Unit, CERN, CH-1211 Gen\`eve 23, Switzerland 
           \and
           S. Moroz \at
           Institut f\"ur Theoretische Physik, Philosophenweg 16, D-69120 Heidelberg, Germany           \and
           R. Schmidt  \at
           Physik Department, Technische Universit\"at M\"unchen, James-Franck-Stra{\ss}e, D-85748 Garching, Germany
}

\date{}

\maketitle

\begin{abstract}
Few-body physics related to the Efimov effect is discussed using the functional renormalization group method. After a short review of renormalization in its modern formulation we apply this formalism to the description of scattering and bound states in few-body systems of identical bosons and distinguishable fermions with two and three components. The Efimov effect leads to a limit cycle in the renormalization group flow. Recently measured three-body loss rates in an ultracold Fermi gas of $^6\textrm{Li}$ atoms are explained within this framework. We also discuss briefly the relation to the many-body physics of the BCS-BEC crossover for two-component fermions and the formation of a trion phase for the case of three species. 

\end{abstract}

\section{Introduction}
\label{intro}
The physics of two or three non-relativistic particles that can scatter of each other or form bound states can be well described by quantum mechanics. Nevertheless, we present here an approach to these problems that is based on a method borrowed from the modern formulation of quantum field theory. Why should this be useful? There are several reasons. First, to better understand few-body physics it is desirable to have several different approaches at hand. Intuitive pictures and quantitative results can be compared and improved. Second, by applying powerful but also involved methods to rather simple problems one can also learn how to improve these methods. Also, by establishing a unified description of different physical phenomena (possibly on very different energy scales) one can hope for fruitful interference between different fields. Some universal aspects might even be transferred directly. Finally, for the field of ultracold quantum gases one has the very concrete advantage that functional renormalization can not only be applied to few-body problems but allows in a very similar way also the description of many-body phenomena.

Our formalism uses the quantum effective action $\Gamma[\psi]$ which is a functional of the field $\psi(t,\vec x)$ representing in our case non-relativistic particles. Formally speaking, the quantum effective action $\Gamma[\psi]$ is the generating functional of one-particle irreducible Feynman diagrams. This means that tree level expressions for scattering amplitudes are exact when the vertices and the propagators are obtained from functional derivatives of $\Gamma[\psi]$. In some sense the effective action $\Gamma[\psi]$ is similar to the microscopic action $S[\psi]$ but with the difference that all quantum and statistical fluctuations are included. Variation of $\Gamma[\psi]$ with respect to $\psi$ yields exact equations. The functional form of $\Gamma[\psi]$ and the connection to scattering and bound state properties will become more clear below. Here we remark that for a generic interacting field theory the calculation of $\Gamma[\psi]$ is non-trivial. A method for this task is provided by functional renormalization.

A particularly interesting feature of few-body physics relevant for experiments with ultracold  quantum gases is universality \cite{BraatenHammer}. Broadly speaking this means that for a large class of microscopic properties (such as the interaction potential between two particles) the macroscopic, low-energy properties are independent of the microscopic details and depend only on a few parameters. In particular if the scattering length is large compared to the effective range of the interaction potential the macroscopic properties can be equivalently calculated from a contact potential (with a vanishing range). In a renormalization group picture universality manifests itself as a (partial) fixed point. A large class of microscopic theories is attracted towards the same RG trajectory during the flow towards the infrared. The low-energy properties of the whole class can be characterized in terms of one representative which is for simplicity chosen to be the contact potential. For this case the microscopic or classical action has a particularly simple form. For a single boson field $\psi$ (as appropriate to describe identical bosons) it reads
\begin{equation}
S = \int d\tau \int d^3 x \; \left\{ \psi^* \left(\partial_\tau - \tfrac{\vec \nabla^2}{2M_\psi}-\mu\right) \psi + \frac{1}{2} \lambda_\psi (\psi^*\psi)^2 \right\}.
\label{eq:microscopicactionbosons}
\end{equation}
The microscopic interaction strength $\lambda_\psi$ will later be related to the s-wave scattering length $a$. We use in Eq.\ \eqref{eq:microscopicactionbosons} a formulation with imaginary time $\tau$ which is integrated from $0$ to $1/T$ in the Matsubara formalism. We also introduced a chemical potential $\mu$. Although we are interested here only in the few-body properties in vacuum, i.e. for vanishing temperature $T=0$ and particle density $n=0$, the generalization to many-body physics at nonvanishing temperature is possible and straightforward in principle by adjusting $\mu$ and $T$. Real-time properties can be obtained from analytic continuation $t=i \tau$. Throughout this paper we will use natural units with $\hbar=k_{B}=1$. In addition we measure energies in units of momentum squared which leads to $2M_{\psi}=1$.

Signatures of universality also arise in few-body physics related to the Efimov effect \cite{Efimov70}. A remarkable feature which underlies the Efimov effect is the presence of a discrete scaling symmetry. Specifically, at the unitarity point, where the scattering length diverges, the solution of the Efimov problem remains invariant (self-similar) under rescaling of length by integer powers of some discrete scaling factor. In the renormalization group language the discrete scaling symmetry is reflected in a renormalization group flow that periodically traverses a closed orbit, a so-called limit cycle \cite{Bedaque}. Limit cycles encode the essence of the Efimov effect and broaden the concept of few-body universality introduced in the previous paragraph. Recent experiments with ultracold atoms (for a summary see \cite{Ferlainorev}) provided the first clear evidence of how Efimov physics is realized in nature.

Since the three-body sector develops a renormalization group cycle scaling at the unitarity point, one must provide a proper ultraviolet initial condition for the three-body interaction, which is dictated by the details of the microscopic physics such as for instance the underlying interatomic potentials. Phrased in RG language, the limit cycle scaling implies that a three-body term $\sim \lambda_3 (\psi^* \psi)^3$ has a complex anomalous dimension and is exactly marginal at unitarity. For these reasons a three-body interaction term must in principle be included in the microscopic action \eqref{eq:microscopicactionbosons}. In the main part of this work we do, however, not concentrate on the direct comparison with particular experiments in which case the microscopical details and correspondingly a three-body term would become of relevance. For the sake of simplicity we tune the three-body parameter $\lambda_3$ to zero which is why such a term is not included in the microscopic action \eqref{eq:microscopicactionbosons}. In this simple setting all three-body correlations are generated during renormalization group evolution. Only in section \ref{sec:6}, where we compare our findings with recent experiments with three-component fermionic $^6\text{Li}$ atoms, the three-body term becomes a part the microscopic description.

This paper is organized as follows. In section \ref{sec:1} we give a short introduction to the functional renormalization method with emphasis on the non-relativistic few-body problem. In section \ref{sec:2} we discuss the problem of two identical bosons with contact interaction as well as the problem of three identical bosons at a broad Feshbach resonance where the scattering length diverges. The subsequent section \ref{sec:4} is dedicated to a system of fermions with two species (e.g. hyperfine states). We discuss how the renormalization group flow equations are modified compared to the case of bosons in the few-body sector and sketch how they account for the BCS-BEC crossover physics at non-zero density. Section \ref{sec:5} contains a discussion of our renormalization group treatment for a system with three fermion species. We assume that the fermions have equal properties (apart from their hyperfine spin) and that there is a common Feshbach resonance for all particles leading to a global $\textrm{SU}(3)$ symmetry. We generalize this model to the case where the resonances occur at different values of the magnetic field in section \ref{sec:6}. In this setup we explain three-body loss rates recently measured in experiments with $^6\text{Li}$ atoms \cite{Ottenstein08,Huckans09}. Finally, we draw our conclusions in section \ref{sec:conclusions}.

\section{Method}
\label{sec:1}

\subsection{Functional renormalization group}\label{sec:frg}

Functional renormalization is a modern realization of Wilson's renormalization group concept which is used in quantum and statistical field theory, especially when dealing with strongly interacting systems (for reviews see \cite{reviews}). The method combines functional methods of quantum field theory with the physically intuitive renormalization group idea. The main motivation for its development was the observation that it is often more useful and transparent to perform the integration of quantum fluctuations in continuous steps rather than doing it at once. The renormalization technique allows to interpolate smoothly between the known microscopic laws and the complicated macroscopic phenomena. In this sense, it bridges the transition from simplicity of microphysics to complexity of macrophysics.

In quantum field theory a particular useful quantity is the effective action $\Gamma$ which is a direct analogue of the classical action functional $S$. It depends on the fields of a given theory and includes all quantum fluctuations.  For a generic interacting field theory the effective action $\Gamma$, however, is difficult to obtain. Functional renormalization provides a practical tool to calculate $\Gamma$ employing the concept of renormalization group.

The central idea of functional renormalization is the introduction of a scale-dependent effective action functional $\Gamma_{k}$ often called average action or flowing action. The average action $\Gamma_k$ interpolates smoothly as a function of the RG sliding scale $k$ between the known microscopic action $S$ and the full quantum effective action $\Gamma$.  The dependence on the sliding scale $k$ is introduced by adding an infrared regulator term to the microscopic action. This regulator term has the purpose to suppress fluctuations below the scale $k$ and it is introduced by defining (for a bosonic field $\tilde\psi$) the $k$-dependent Schwinger functional $W_{k}[J]$
\begin{equation}
e^{W_{k}[J]} = \int \mathcal{D} \tilde\psi \; e^{-S[\tilde \psi]-\Delta S_{k}[\tilde \psi]+\int_{x} J \tilde \psi}.
\label{kdepSchwingerfunct}
\end{equation}
The difference to the standard QFT Schwinger functional $W[J]$ is the additional term $\Delta S_{k}[\tilde \psi]$ in the exponential in Eq. \eqref{kdepSchwingerfunct}. $\Delta S_{k}[\tilde \psi]$ is quadratic in the fields and in momentum space it reads
\begin{equation}
\Delta S_{k}[\tilde\psi] = \int_{Q} \tilde\psi^{*}(Q) R_{k}(Q) \tilde\psi(Q).
\end{equation}
We use the abbreviations $Q=(q_0, \vec q)$ and $\int_Q = \int_{q_0} \int_{\vec q}$ with $\int_{q_0} = \tfrac{1}{2\pi} \int_{-\infty}^\infty dq_0$ and $\int_{\vec q} = \tfrac{1}{(2\pi)^3} \int d^3 q$. The infrared regulator function $R_k(Q)$ is not fixed in the unique way, but must satisfy three important conditions. First $R_k(Q)$ should suppress infrared modes below the scale $k$, i.e.
\beq \label{rc1}
R_k(0)\approx k^2.
\eeq
This adds effectively an additional masslike term to the particle suppressing its propagation if its momentum is below the RG scale $k$. Second, the regulator must vanish in the infrared
\beq \label{rc2}
\lim\limits_{k\to 0} R_k(Q) = 0.
\eeq
With this condition it is guaranteed that $W_k$ equals the standard, exact Schwinger functional $W$ in the infrared, $W=W_{k=0}$. Finally, we should recover the classical action in the ultraviolet and thus we demand
\beq \label{rc3}
\lim\limits_{k\to \infty} R_k(Q) = \infty.
\eeq
Apart from these requirements, the regulator can be chosen arbitrarily. Nevertheless, it is recommended to pick the regulator carefully depending on a concrete physical problem. Moreover, one should better choose a regulator that respects as many symmetries of the studied problem as possible.
For our purpose it is convenient to choose $R_k(Q)=R_k(\vec q)$ independent of the frequency argument $q_0$. Often, $R_k(\vec q)$ is chosen to decay for large $\vec q^2 \gg k^2$. As we have seen, in Eq.\ \eqref{kdepSchwingerfunct} the regulator $R_k$ suppresses fluctuation modes with momenta $\vec q^2\lesssim k^2$ by giving them a large ``mass'' or ``gap'', while high momentum modes are not (or only mildly) affected. Thus, $W_{k}$ includes all fluctuations with momenta $\vec q^2\gtrsim k^2$. 

The flowing action $\Gamma_k$ is now obtained by subtracting from the Legendre transform of the Schwinger functional,
\begin{equation}
\tilde \Gamma_k[\psi] = \int J \psi - W_k[J]
\end{equation}
with $\psi = \delta W_k[J]/\delta J$, the cutoff term
\begin{equation}
\Gamma_k[\psi] = \tilde \Gamma_k[\psi] - \Delta S_k [\psi]. 
\end{equation}

The reason why the flowing action $\Gamma_k[\psi]$ is the central object to study is that it obeys the elegant and exact functional flow equation
\begin{eqnarray}\label{M1}
 \partial_k \Gamma_k[\psi] &=& \frac{1}{2} \text{STr} \,
 \partial_k R_k \, (\Gamma^{(2)}_k[\psi] + R_k)^{-1},
\end{eqnarray}
derived by C. Wetterich in 1993 \cite{Wetterich}. Despite its compact form, the flow equation governs the full quantum dynamics. In Eq. \eqref{M1} $\partial_k$ denotes a derivative with respect to the sliding scale $k$ at fixed values of the fields.
The functional differential equation for $\Gamma_{k}[\psi]$ must be supplemented with the initial condition $\Gamma_{k\to\Lambda}=S$, where the ``classical action'' $S[\psi]$ describes the physics at the microscopic ultraviolet scale $k=\Lambda$. Importantly, in the infrared limit $k\to 0$ the full quantum effective action $\Gamma[\psi]=\Gamma_{k\to 0}[\psi]$ is obtained. In the Wetterich equation $\text{STr}$ denotes a supertrace operation which sums over momenta, frequencies, internal indices, and different fields (taking bosons with a plus and fermions with a minus sign). 

It is important to note that the exact flow equation for $\Gamma_k[\psi]$ has a one-loop structure. This is a significant simplification compared for example to perturbation theory, where all multi-loop Feynman diagrams must be included. Note however, that Eq.\ \eqref{M1} is a nonlinear functional differential equation. Indeed, on the right-hand side in the denominator appears the second functional derivative $\Gamma_k^{(2)}[\psi]$. Due to that it is usually not possible to find closed solutions to Eq.\ \eqref{M1} in an analytical form. 

\begin{figure}
\begin{center}
\includegraphics*[height=5cm,angle=0]{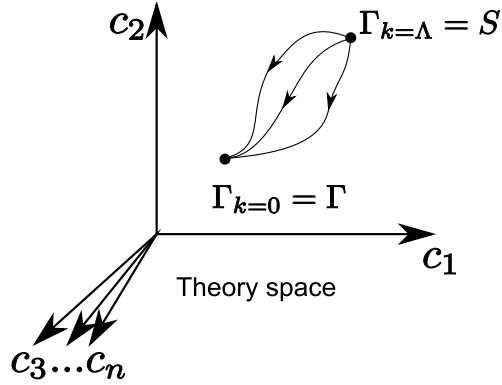}
\end{center}
\caption{Renormalization group flow in the theory space of all possible couplings allowed by symmetries. Different trajectories correspond to different choices of the infrared regulator $R_k$.}
\label{fig1}
\end{figure}

To make the implications of \eqref{M1} more transparent we write the flowing action as an expansion in some complete set of operators
\begin{equation}
\Gamma_k[\psi] = \sum_{n=0}^\infty c_n \; {\cal O}[\psi].
\end{equation}
The renormalization group evolution of $\Gamma_k$ traces a trajectory in the theory space, which is the infinite-dimensional space of all possible couplings $\{c_{n} \}$ allowed by the symmetries of the problem. As schematically shown in Fig. \ref{fig1}, at the  microscopic ultraviolet scale $k=\Lambda$ one starts with the initial condition $\Gamma_{k=\Lambda}=S$.
As the sliding scale $k$ is lowered, the flowing action $\Gamma_k$ evolves in theory space according to the functional flow equation \eqref{M1}. Accordingly, different couplings $\{c_n \}$ become running or flowing under the renormalization group evolution. The choice of the regulator $R_k$ is not unique, which introduces some scheme dependence into the renormalization group flow. For this reason, different choices of the regulator $R_k$ correspond to the different paths in Fig. \ref{fig1}. At the infrared scale $k=0$, however, the full effective action $\Gamma_{k=0}=\Gamma$ is recovered for every choice of the cutoff $R_k$, and all trajectories meet at the same point in the theory space.

\subsection{Non-relativistic few-body problem} \label{nfp}
As was mentioned above we use here a field-theoretic formalism with which also many-body problems in thermal and chemical equilibrium can be addressed. In this setup the properties of a few particles such as scattering amplitudes and binding energies are encoded in the $n$-point correlation functions in vacuum. The projection of the effective action $\Gamma$ onto the vacuum state was developed in \cite{Diehl:2005ae,DKS}. It amounts in cooling the system ($T\to 0$) and making it dilute ($n\to 0$) in such a way, that the temperature $T$ always stays above its critical value in order to avoid many-body effects (e.g., Bose-Einstein condensation). The chemical potentials of different particles in a given problem are related to their binding energies measured with respect to some reference energy level. As usual in non-relativistic physics the absolute energy scale can be shifted, and it is often convenient for us to fix it such that the lowest energy excitation (lowest energy $n$-particle bound state) has a vanishing energy. 

All few-body systems discussed here are translation invariant, i.e. no trapping potential is considered. Moreover, they are invariant under space-time Galilei transformations, which restricts the form of the non-relativistic propagator of a field $\psi$ to be a function of the combination  $\partial_{\tau}-\vec\nabla^2$ of temporal and spatial derivatives (we recall that in our units $2M_{\psi}=1$). In addition, all systems possess an internal $\textrm{U}(1)$ symmetry associated with particle number conservation.

As was discussed in detail in \cite{DKS,MFSW,DFGPW,SFDoctoralThesis}, numerous mathematical simplifications appear in few-body problems compared with more complicated many-body problems. For instance, all Feynman diagrams with loop lines pointing in the same direction necessarily vanish. This simplification originates from the fact that there are only particles, but no antiparticles or holes as excitations of the non-relativistic vacuum.
As a result, a remarkable hierarchy of RG flow equations arises in few-body physics. The flow equations of the n-body sector are not influenced by  the higher-body sectors.\footnote{The $n$-body sector is defined here as the set of one-particle irreducible $2n$-point functions expressed in terms of the elementary $\psi$-fields. For a detailed discussion of the few-body hierarchy as well as a general proof of this feature see \cite{SFDoctoralThesis}.} One can solve the two-particle problem without any information about the three-particle problem and so on. We note that the presence of this hierarchy makes the systematic vertex expansion, discussed in Sec. \ref{ve}, very useful for the treatment of few-body problems. The generally applicable strategy thus is to solve $n$-body problems subsequently by starting with the lower (one-particle, two-particle) sectors and proceeding to the higher ones.

\subsection{Truncations} \label{truncations}
In most cases of interest the flow equation \eqref{M1} can only be solved approximately. Usually some type of expansion of $\Gamma_{k}$ is first performed, which is then truncated at finite order leading to a finite system of ordinary differential equations. The choice of a suitable expansion scheme should be physically motivated and depends on the given problem. The expansions do not necessarily involve a small parameter (like an interaction coupling constant or $1/N$) and thus they are, in general, of nonperturbative nature. Here we describe two different expansion schemes that we used for studying few-body quantum problems with functional renormalization.

\subsubsection{Derivative expansion} \label{de}
A particularly simple truncation scheme is the derivative expansion. One expands the flowing action in terms with increasing number of spatial and temporal derivatives.\footnote{Due to the non-relativistic nature of the problem we can assign to the derivatives the scaling dimensions $[\vec{\nabla}]=1$ and $[\partial_{\tau}]=2$ which corresponds to the dynamical critical exponent $z=2$.} As an example consider bosons with the microscopic action given in Eq. \eqref{eq:microscopicactionbosons}. In the derivative expansion the ansatz for the flowing action is
\begin{equation}
\begin{split}
\Gamma_k[\psi] = \int_{\tau,\vec x} {\bigg \{ } & U_k(\psi^*\psi) + Z_1 \psi^* \partial_\tau \psi + Z_2 \psi^*(-\vec \nabla^2) \psi\\
& + V_1 \psi^* (-\partial_\tau^2 ) \psi + V_2 \psi^*(\partial_\tau \vec \nabla^2) \psi + V_3 \psi^*(-\vec  \nabla^4)\psi+\dots {\bigg \} }.
\end{split}
\label{eq:derivativeexpansion}
\end{equation}
Similar to the effective potential $U_k(\psi^*\psi)$, being a function of $\psi^*\psi$, the coefficients $Z_i$ and $V_i$ are functions of the scale parameter $k$ and depend in principle also on the $\textrm{U}(1)$ invariant combination $\psi^*\psi$. Besides higher order derivatives also terms of the form $\psi^*\psi (-\vec \nabla^2)^n \psi^*\psi$ will appear in Eq. \eqref{eq:derivativeexpansion}. 

The derivative expansion is particularly useful for the investigation of many-body problems \cite{reviews}. In problems where bound state formation plays a role one expects that the derivative expansion in the form \eqref{eq:derivativeexpansion} is not particularly useful. The reason is that one expects in this case a pole in the effective four-point correlation function. Extracting this object from Eq. \eqref{eq:derivativeexpansion} by taking the appropriate functional derivatives, one finds that it is given by a combination of terms with different frequency and momentum dependence. However, all these terms are regular and any finite number of them cannot account for a bound state pole.

For this reason, instead of working with the microscopic action in Eq. \eqref{eq:microscopicactionbosons} it is sometimes useful to perform a Hubbard-Stratonovich transformation. One adds a term that is quadratic in the complex auxiliary field $\phi$. The action becomes
\begin{equation}
\begin{split}
S = \int_{\tau, \vec x} {\bigg \{ } & \psi^*(\partial_\tau - \vec \nabla^2-\mu) \psi + \tfrac{1}{2} \lambda_\psi (\psi^*\psi)^2\\
& + \left(\phi^* - \psi^* \psi^* \tfrac{h}{m_\phi^2}\right)  m_\phi^2 \left(\phi - \tfrac{h}{m_\phi^2}\psi \psi \right).
\end{split}
\label{eq:1}
\end{equation}
Since the action for $\phi$ is Gaussian, the field $\phi$ can be easily integrated out. Hence, the action \eqref{eq:1} is equivalent to the original action \eqref{eq:microscopicactionbosons}, and under the condition 
\begin{equation}
\frac{h^2}{m_\phi^2}+ \tfrac{1}{2}\lambda_\psi = 0,
\label{eq:2}
\end{equation}
the interaction term $\sim \lambda_\psi$ is cancelled. The resulting theory is of the Yukawa type
\begin{equation}
\begin{split}
S = \int_{\tau, \vec x} {\bigg \{} & \psi^*(\partial_\tau - \vec \nabla^2-\mu) \psi - h (\phi^*\psi \psi + \psi^*\psi^*\phi)
 + \phi^* m_\phi^2 \phi {\bigg \}}.
\end{split}
\label{eq:3}
\end{equation}
It is possible to investigate the theory in this form using a derivative expansion for the original field $\psi$ as well as the composite auxiliary (dimer) field $\phi$. This will be described in more detail below.

For some purposes it is useful to do the Hubbard-Stratonovich transformation not at a single scale but in a $k$-dependent way. This is advantageous in particular when the interaction parameter that develops a bound-state pole is due to fluctuation effects and emerges from the functional renormalization flow. An exact flow equation that realizes such a scale-dependent Hubbard-Stratonovich transformation has been derived \cite{Bosonization,FWCompositeOperators}. For a discussion of this approach to the treatment of bound states in the context of non-relativistic physics see ref. \cite{SFBoundStates}.

\subsubsection{Vertex expansion} \label{ve}
As we advocated in Sec. \ref{nfp}, a systematic vertex expansion of $\Gamma_k$ is especially suitable for studies of few-body problems. The vertex expansion is an expansion in powers of fields, i.e. schematically
\beq \label{ve1} \Gamma_{k}=\sum_{n=2}^{\infty}\Gamma_{k}(n)=\Gamma_{k}(2)+\Gamma_{k}(3)+\Gamma_{k}(4)+..., \eeq
where $n$ represents the number of fields in the monomial $\Gamma_k(n)$. We dropped $\Gamma_k(0)$ in \eqref{ve1} which encodes the ground state energy of the non-relativistic (Fock) vacuum. The term linear in fields is absent due to $\textrm{U}(1)$ symmetry. It is important to stress that, even if absent in the microscopic action $S=\Gamma_{k=\Lambda}$, the vertices $\Gamma_k(n)$ become in general nonlocal in position space during the RG flow. Assuming translation invariance it is convenient to Fourier transform to momentum space, where the monomials $\Gamma_k(n)$ are functions of $n-1$ kinematically independent momenta and energies. At the infrared scale ($k=0$) this kinematic dependence of $\Gamma_k$ directly translates into energy and momentum dependence of scattering and bound state properties of the system. 

To make this construction transparent we consider  the example of bosons described by the microscopic action $S$ in Eq. \eqref{eq:microscopicactionbosons}. The term $\Gamma_k(3)$ vanishes in this case, and the vertex expansion in Eq.\ \eqref{ve1} starts with the terms 
\beq \label{ve2}
\begin{split}
\Gamma_{k}(2)&=\int_Q \psi^* (Q)P_{\psi}(Q)\psi(Q) \\
 \Gamma_{k}(4)&=\frac{1}{2}\int_{Q_1,\dots Q_4 } \lambda_{\psi}(Q_1, Q_2, Q_3) \psi^* (Q_1)\psi^*(Q_2) \psi(Q_3) \psi(Q_1+Q_2-Q_3),
\end{split}
\eeq
where $P_{\psi}(Q)$ is the scale-dependent inverse propagator and $\lambda_{\psi}(Q_1, Q_2, Q_3)$ is the nonlocal scale-dependent four-particle vertex. Due to the $\textrm{U}(1)$ particle number symmetry only terms with equal number of $\psi$ and $\psi^*$ can appear in the vertex expansion.


\section{Identical bosons}
\label{sec:2}
In this section we investigate the few-body physics of identical bosons interacting via a contact interaction. The hierarchy of flow equations introduced in section \ref{nfp} turns out to be extremely useful as it allows us to tackle few-body problems step by step. Completely detached from higher n-body sectors we can first analyze the flow equations describing a single particle, then proceed to the two-body sector followed by the three-body system. For solving the $n$-body problem we only need the solution of the $m$-body systems with $m<n$.

\subsection{Two-body problem} \label{two-bodyproblem}

The interacting bosons are governed by the microscopic action in Eq. \eqref{eq:microscopicactionbosons}. In the following it will be useful to switch to the formulation after the  Hubbard-Stratonovich transformation in Eq.\ \eqref{eq:3}.

In order to apply the functional renormalization group equation \eqref{M1} we first have to choose a truncation for the effective flowing action $\Gamma_k$. We choose a vertex expansion as introduced in Eq. \eqref{ve1}, which, up to third order in fields, reads in momentum space
\begin{eqnarray}
\label{rseq:vebosonstwobody}
\Gamma_k(2)&=&\int_Q \psi^*(Q)(i q_0+\bold q^2-\mu_\psi)\psi(Q) + \int_Q \phi^*(Q)P_\phi(Q)\phi(Q),\nonumber \\
\Gamma_k(3)&=&\frac{h}{2}\int_{Q_1,Q_2,Q_3}[ \phi^*(Q_1)\psi(Q_2)\psi(Q_3)+\phi(Q_1)\psi^*(Q_2)\psi^*(Q_3)] \delta(Q_1-Q_2-Q_3).
\end{eqnarray}
We stress that $\Gamma_k(2)$ and $\Gamma_k(3)$ do not have the most general form compatible with Galilean and global $\textrm{U}(1)$ particle number symmetry. This would include a general inverse atom propagator $P_\psi(Q)$ and a momentum dependent Yukawa coupling $h(Q_1,Q_2)$. However, using the special properties of the vacuum one can show that neither $P_\psi(Q)$ nor $h(Q_1,Q_2)$ get renormalized. Indeed, the condition of vanishing density implies that the poles of the propagators with respect to the frequency argument $q_0$ are restricted to the upper complex half-plane. Therefore diagrams with all internal lines pointing in the same direction (we call these diagrams `cyclic') have integrands  whose poles lie in the upper complex frequency half-plane. By closing the integration contour in the lower half-plane one finds that the corresponding expressions vanish.

\begin{figure}[t!]
\begin{minipage}[c]{0.5\textwidth}
\centering
\includegraphics[width=0.6\textwidth]{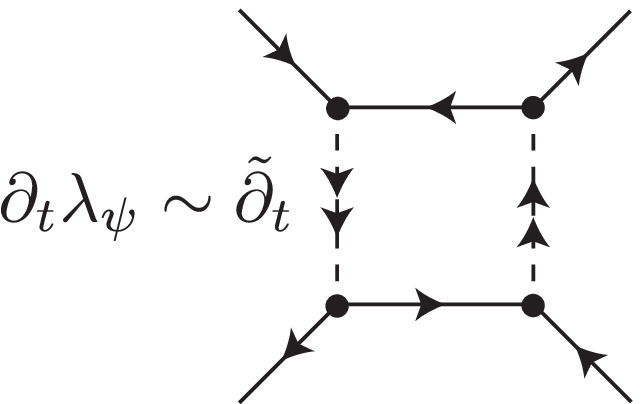}
\end{minipage}%
\begin{minipage}[c]{0.5\textwidth}
\centering 
\includegraphics[width=0.85\textwidth]{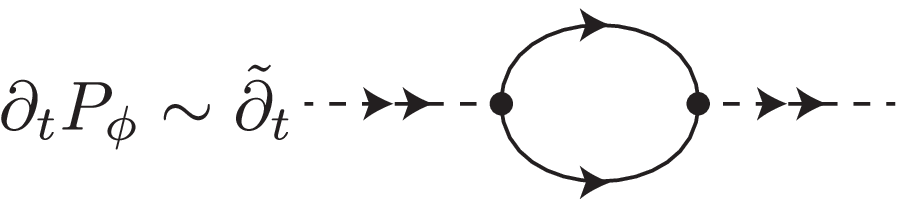}
\end{minipage}
\caption{The beta functions of the coupling $\lambda_\psi$ (left) and the dimer propagator $P_\phi$ (right) depicted schematically as one-loop Feynman diagrams. The arrows in Feynman diagrams correspond to flow of particle number associated with $\textrm{U}(1)$ symmetry. }
\label{fig:twobodyflows}
\end{figure}

Let us investigate now the renormalization of the two-body sector. Here, we have to consider the couplings $h$, $\lambda_\psi$, and the inverse dimer propagator $P_\phi$. At the UV scale, $k=\Lambda$, the contact coupling $\lambda_\psi$ vanishes because it has been replaced by the dimer exchange interaction in the Hubbard-Stratonovich transformation. Yet there is the box diagram corresponding to the $\beta$-function $\partial_t\lambda_\psi$, schematically shown in Fig. \ref{fig:twobodyflows} (left). This diagram potentially regenerates $\lambda_\psi$ during the RG flow. It is however cyclic and thus one has $\lambda_\psi\equiv0$ during the whole RG flow. The only possible diagram renormalizing the Yukawa coupling $h$ contains the vertex $\lambda_\psi$, which is zero. For this reason also $h$ is not renormalized, and one is left with the flow equation for the inverse dimer propagator $P_\phi(Q)$. For this reason, the truncation \eqref{rseq:vebosonstwobody} is complete to third order in fields.

The diagram renormalizing $P_\phi$ is shown in Fig. \ref{fig:twobodyflows} (right) and can be written as follows:\footnote{Here we use the logarithmic, dimensionless RG scale $t=\ln\frac{k}{\Lambda}$ such that $\partial_t=k\partial_k$.}
 \begin{equation}\label{eq:twobodydimer1}
\partial_{t}P_{\phi}(Q)=-\frac{2}{3+p}\int\limits_{L}\tilde{\partial}_{t}\frac{h^2}{(P_{\psi}(L)+R_{\psi}(L))(P_{\psi}(Q-L)+R_{\psi}(Q-L))},
\end{equation}
with $p=1$ for bosons $\psi$,\footnote{The scattering physics of two distinguishable particles such as two fermions with opposite spin can be solved analogously with $p=-1$ \cite{MFSW}.}
 \begin{equation} \label{atomprop}
P_{\psi}(Q)=i q_0+\mathbf{q}^2-\mu_{\psi}
\end{equation}
and $R_\psi(Q)$ denotes the atom regulator. In Eq.\ \eqref{eq:twobodydimer1}, the symbol $\tilde\partial_t$ represents a formal $t$ derivative that hits only the cutoff $R_\psi(Q)$. For the choice of $R_\psi$ we follow \cite{DKS,MFSW} and choose a frequency and momentum independent cutoff of the simple form $R_\psi=k^2$. It turns out that this masslike regulator allows a relatively simple, analytical integration of the flow equation \eqref{eq:twobodydimer1}. The two-body problem has also been solved analytically using the so called optimized Litim cutoff by Birse \cite{Birse08}. In contrast to the $k^2$ regulator, the Litim regulator breaks Galilean invariance such that symmetry restoring counterterms have to be inserted in the UV.

First we perform the frequency loop integration in Eq. \eqref{eq:twobodydimer1} with the help of the residue theorem:
\begin{eqnarray} \label{eq:twobodystep1}
\partial_{t}P_{\phi}(Q)&=&-\frac{2}{3+p}\int \frac{d^{3}l}{(2\pi)^3}\tilde{\partial}_{t}\frac{h^2}{i q_0+\mathbf{l}^2+(\mathbf{l}-\mathbf{q})^2-2\mu_{\psi}+2R_{\psi}} \nonumber \\
&=&-\frac{2h^2}{3+p}\partial_{t}\int \frac{d^{3}l}{(2\pi)^3}\frac{1}{i q_0+\mathbf{l}^2+(\mathbf{l}-\mathbf{q})^2-2\mu_{\psi}+2R_{\psi}}.
\end{eqnarray}
In the second line we replaced $\tilde\partial_t$ with $\partial_t$. The tilde indicates that the derivative acts only upon the $k$-dependence of the regulator. We are allowed to drop the tilde as neither the Yukawa coupling $h$ nor the atom propagator $P_\psi$ is renormalized such that only the regulator term exhibits a $k$-dependence. This permits to perform the $t$ integration of the flow equation from the UV scale $t_\textrm{UV}=0$ to the IR scale $t_\textrm{IR}=-\infty$. Using the specific values of the regulator $R_\psi(t_{\textrm{IR}})=0$ and $R_\psi(t_{\textrm{UV}})=\Lambda^2$ we obtain
\begin{eqnarray} \label{eq:twobodydimer2}
&&P^\textrm{IR}_{\phi}(Q)-P^\textrm{UV}_{\phi}(Q)\nonumber\\
&&=-\frac{2h^2}{3+p}\int \frac{d^{3}l}{(2\pi)^{3}}\left( \frac{1}{iq_0+\mathbf{l}^2+(\mathbf{l}-\mathbf{q})^2-2\mu_{\psi}}-\frac{1}{iq_0+\mathbf{l}^2+(\mathbf{l}-\mathbf{q})^2-2\mu_{\psi}+2\Lambda^{2}} \right) \nonumber \\
&&=-\frac{h^2}{3+p}\int\frac{dl}{2\pi^2}\left( \frac{l^2}{l^2+\big(\frac{iq_0}{2}+\frac{\mathbf{q}^2}{4}-\mu_{\psi}\big)} -\frac{l^2}{l^2+\big(\frac{iq_0}{2}+\frac{\mathbf{q}^2}{4}-\mu_{\psi}+\Lambda^{2}\big)} \right) \nonumber \\
&&=-\frac{h^2}{4\pi (3+p)}\big(\Lambda-\sqrt{\frac{iq_0}{2}+\frac{\mathbf{q}^2}{4}-\mu_{\psi}} \big).
\end{eqnarray}
In the first line we get rid of the angular integration by substituting $\bold l\to \bold l - \bold q/2$ and the last identity assumes  $\Lambda>> |\mu_{\psi}|, |\mathbf{q}|, |q_0|$.

At this point we have to fix the initial condition for $P_\phi^\textrm{UV}$ at $k=\Lambda$ in order to obtain the physical inverse propagator $P_\phi^\textrm{IR}$ at $k=0$. The momentum and frequency independent part of the inverse dimer propagator at the microscopic scale can be written as
\begin{equation}
P_\phi^\textrm{UV}= \delta\nu+\nu(a).
\end{equation}
The physical properties determined by the inverse infrared propagator $P_\phi^\textrm{IR}$, should be independent of the cutoff scale $\Lambda$. From Eq. \eqref{eq:twobodydimer2} we see, that this condition determines the term $\delta\nu$,
\begin{equation}
\delta\nu=\frac{h^2}{4\pi (3+p)}\Lambda.
\end{equation}
Furthermore, as the composite dimer $\phi$ mediates the interaction between atoms, $P_\phi^\textrm{IR}$ has to correctly describe the scattering physics characterized by the s-wave, zero-range scattering amplitude
\begin{equation}\label{eq:qmscattam}
f(q)=\frac{1}{-a^{-1}-i q},
\end{equation}
where $E=2\bold q^2$ is the total energy of the two scattered atoms with momenta $\pm\bold q$ respectively. The scattering amplitude is in turn connected to $P_\phi^\textrm{IR}$. A simple tree level analysis shows
\begin{equation}
 f(0)=- a=\frac{1}{16\pi} \frac{h^2}{P^\textrm{IR}_\phi(0,0,\mu_\psi=0)},
\end{equation}
where the dimer propagator is evaluated for vanishing frequency and momentum. At this point it is important to note, that when describing the s-wave scattering of atoms one deals with atoms at the energy threshold in the in- and outgoing states. For this reason it is necessary to work at vanishing atom energy, $\mu_\psi=0$. Taking this into account we are in the position to determine $\nu(a)$. Evaluating Eq. \eqref{eq:twobodydimer2} for vanishing momentum, energy and atom gap $\mu_\psi$ we find
\begin{equation}
\nu(a)=-\frac{h^2}{4\pi(3+p)}a^{-1}.
\end{equation} 
We finally arrive at the solution of the two-body problem encoded in the exact infrared inverse dimer propagator
\begin{equation} \label{eq:exactdimerirprop}
P_{\phi}(Q)\equiv P_{\phi, k=0}(Q)=\frac{h^2}{4\pi(3+p)}\Big(-a^{-1}+\sqrt{\frac{i q_0}{2}+\frac{\mathbf{q}^2}{4}-\mu_{\psi}} \Big).
\end{equation}
This propagator also recovers -- after performing the Wick rotation to real energies -- the exact (s-wave) zero-range scattering amplitude in Eq. \eqref{eq:qmscattam}.

Quantum mechanics predicts the existence of a universal, weakly bound dimer state for two particles with an interaction characterized by a large, positive scattering length $a$. Its binding energy can be extracted from a pole of the T-matrix at negative energy. In our approach the T-matrix is given by
\begin{equation}
\tau(E)=-\frac{h^2}{P_\phi^\textrm{IR}(0,0,\mu_\psi=E/2)}
\end{equation}
and finding its pole therefore corresponds to solving $P_\phi^\textrm{IR}(0,0,\mu_\psi)=0$ for the atom gap $-\mu_\psi>0$. This leads to
\begin{equation}
a=\frac{1}{\sqrt{-\mu_\psi}}.
\end{equation}
As the dimer consists of two atoms we find for its binding energy
\begin{equation} \label{eq:dimerbindingenergy1}
E_{\textrm{D}}=2\mu_{\psi}=-\frac{2}{a^{2}}, \qquad E_\textrm{D}=-\frac{\hbar^2}{m a^{2}}.
\end{equation}
The second equation is expressed in conventional units and is the well-known universal relation for the binding energy of the shallow dimer. The determination of $\mu_\psi$ takes care of the correct vacuum limit. The atom gap $\mu_\psi<0$ is fixed to the value where the dimer becomes gapless in the infrared. This reflects that for $a>0$ the dimer is the two-body ground state  of the system. 

\subsection{Three-body problem} \label{three-body problem}
In order to treat the three-body system of identical bosons within the vertex expansion scheme, we extend our truncation \eqref{rseq:vebosonstwobody} by the term
\begin{equation} \label{three1}
\Gamma_k(4)=-\int\limits_{Q_1,...Q_4}  \lambda_{3}(Q_{1},Q_{2},Q_{3})\phi(Q_{1})\psi(Q_{2})\phi^{*}(Q_{3})\psi^{*}(Q_{4})\delta(Q_{1}+Q_{2}-Q_{3}-Q_{4})
\end{equation}
with $\lambda_{3}(Q_{1},Q_{2},Q_{3})$ denoting the atom-dimer vertex. For simplicity we restrict our discussion to unitarity, $a^{-1}=0$, and demonstrate how Efimov physics emerges from the renormalization group flow equation for the atom-dimer coupling $\lambda_{3}(Q_{1},Q_{2},Q_{3})$.

\begin{figure}[t]
\centering
\includegraphics[width=100pt, height=100pt]{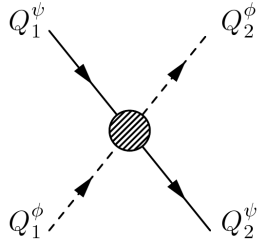}
\caption{\small{Kinematics of the vertex $\lambda_{3}(Q_{1}^{\psi},Q_{2}^{\psi},E)$ in the center-of-mass frame. The atoms and dimers have momenta $Q_{1}^{\psi}=(E_{\psi 1},\mathbf{q}_{1})$ , $Q_{1}^{\phi}=(-E_{\psi 1}+E,-\mathbf{q}_{1})$ and $Q_{2}^{\psi}=(E_{\psi 2},\mathbf{q}_{2})$, $Q_{2}^{\phi}=(-E_{\psi 2}+E,-\mathbf{q}_{2})$.
}}
\label{fig3}
\end{figure}

First consider the kinematics of the atom-dimer scattering. The 1PI vertex $\lambda_{3}(Q_{1},Q_{2},Q_{3})$ depends on three independent energies and spatial momenta. In the following we work in the center-of-mass frame and take the incoming atom and dimer to have momenta $\mathbf{q}_{1}$ and $-\mathbf{q}_{1}$, and energies $E_{\psi1}$ and $E-E_{\psi1}$, while the outgoing particles have momenta $\mathbf{q}_{2}$ and $-\mathbf{q}_{2}$ and energies $E_{\psi2}$ and $E-E_{\psi2}$. In the following the atom-dimer vertex will be denoted by $\lambda_{3}(Q_{1}^{\psi},Q_{2}^{\psi},E)$. The kinematics is summarized in Fig. \ref{fig3}. 

The closed exact solution of the two-body problem, presented in the previous subsection, provides a simple strategy for the derivation of the renormalization group flow equation for the atom-dimer vertex. In principle, one can introduce completely independent  regulators $R_{\psi}$ and $R_{\phi}$ for the atoms $\psi$ and dimers $\phi$. Nevertheless, since the presence of the cutoff $R_{\phi}$ does not alter the solution of the two-body problem, we choose for the atoms the Galilei-invariant momentum independent cutoff $R_{\psi}$, used in subsection \ref{two-bodyproblem}, and lower it first from $R_{\psi}=\Lambda^2$ to zero while keeping $R_{\phi}$ fixed. As the result of this step the dimer inverse propagator is renormalized according to Eq. \eqref{eq:exactdimerirprop}. This also induces interactions between dimers, as for example a term $\sim (\phi^{*}\phi)^2$. However, these interactions belong to the four-body and higher sectors. By virtue of the vacuum hierarchy, they do not influence the flow of $\lambda_3$. Subsequently in the second step, only the dimer cutoff is present and we lower $R_{\phi}$ from its UV value to zero.  In the second step $P_{\phi}$ and $P_{\psi}$ are fixed according to Eqs. (\ref{eq:exactdimerirprop}) and (\ref{atomprop}). In this subsection we use a sharp cutoff for the dimer field,
\beq \label{th3a}
R_{\phi}(Q,k)=P_{\phi}(Q)\left(\frac{1}{\theta(|\vec q|-k)}-1 \right).
\eeq
The special feature of this cutoff is that the regularized dimer propagator takes the simple form
\beq \label{th3b}
\frac{1}{P_{\phi}(Q)+R_{\phi}(Q,k)}=\theta(|\vec q|-k)\frac{1}{P_{\phi}(Q)}.
\eeq

\begin{figure}[t]
\centering
\includegraphics[width=140mm]{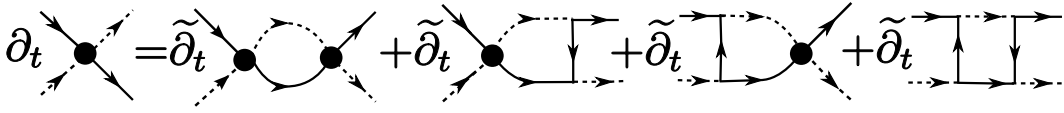}
\caption{\small{Graphical representation of the flow equation for the atom-dimer vertex $\lambda_{3}$. Full lines denote atoms $\psi$ and dashed lines dimers $\phi$. The shaded circle denotes the atom-dimer vertex $\lambda_{3}$.
}}
\label{fig4}
\end{figure}

Following the procedure described above we derive the flow equation for the atom-dimer vertex $\lambda_{3}(Q_{1}^{\psi},Q_{2}^{\psi};E)$, which is given by
\begin{eqnarray}\label{th4}
\partial_t \lambda_{3}(Q^{\psi}_1,Q^{\psi}_2; E) &=&   -\int\limits_L \frac{k\delta(|\mathbf{l}|-k)}{P_{\psi}(L)  P_{\phi}(-L+Q)} \Bigg[C \lambda_{3}(Q_1^{\psi},L;E) \lambda_{3}(L,Q_2^{\psi} ;E) \\\nonumber
&&\qquad \qquad + \frac{B}{2}\Big\{  \frac{h^{2}}{P_{\psi}(-L + Q_1^{\phi})} \lambda_{3}(L,Q_2^{\psi};E)
 +  \lambda_{3}(Q_1^{\psi},L;E) \frac{h^2}{P_{\psi}(-L + Q_2^{\phi})}\Big\}\\\nonumber
&&\qquad\qquad + A\frac{h^2}{P_{\psi}(-L + Q_1^{\phi})}\,\frac{h^2}{P_{\psi}(-L + Q_2^{\phi})}\Bigg],
\end{eqnarray}
where $Q=Q_1^{\phi}+Q_1^{\psi}=(E,\mathbf{0})$, $A=1$, $B=2$ and $C=1$. The graphical representation of this flow equation in terms of one-loop Feynman diagrams can be found in Fig. \ref{fig4}.

Fortunately, at low energies and momenta the flow equation can be considerably simplified (for details see \cite{MFSW}). After introducing the angle-averaged 1PI atom-dimer vertex
\beq \label{th5}
\lambda_{3}(q_{1},q_{2},E)\equiv\frac{1}{2h^2}\int^{1}_{-1}d(\cos\theta)\lambda_{3}(\mathbf{q}_{1},\mathbf{q}_{2};E), \qquad \cos(\theta)=\frac{\mathbf{q}_{1}\cdot \mathbf{q}_{2}}{|\mathbf{q}_{1}||\mathbf{q}_{2}|}
\eeq
and transforming to the real-time formalism, the flow equation reads
\begin{eqnarray} \label{th6}
\partial_{t}\lambda_{3}(q_1,q_2,E)&=&-\frac{2(3+p)}{\pi}\frac{k^3}{\sqrt{\frac{3k^2}{4}-\frac{E}{2}-i\epsilon}} \Bigg[ C \lambda_{3}(q_{1},k,E)\lambda_{3}(k,q_{2},E)+  \nonumber \\
&& \frac{B}{2}\left\{ \lambda_{3}(q_{1},k,E)G(k,q_{2})+G(q_{1},k)\lambda_{3}(k,q_{2},E) \right\}+A G(q_{1},k)G(k,q_{2}) \Bigg],
\end{eqnarray}
where we remind that for bosons $p=1$, and the symmetric function $G(q_{1},q_{2})$ is defined by
\beq \label{th7}
G(q_{1},q_{2})=\frac{1}{4q_{1}q_{2}}\log\frac{q_{1}^2+q_{2}^2+q_{1}q_{2}-\frac{E}{2}-i\epsilon}{q_{1}^2+q_{2}^2-q_{1}q_{2}-\frac{E}{2}-i\epsilon}.
\eeq
The infinitesimally positive $i\epsilon$ term arises from the inverse Wick rotation and makes both Eqs. \eqref{th6} and \eqref{th7} well-defined.
Note that Eq. \eqref{th6} is independent of the Yukawa coupling $h$ and thus is well-defined in the broad resonance limit $h\to\infty$.

The flow equation \eqref{th6}, obtained above, is quite complicated. In order to gain physical understanding, we first employ a crude, but simple and intuitive pointlike approximation. In this approximation the 1PI vertex $\lambda_{3}(q_1,q_2,E)$ is replaced by a momentum and energy independent coupling $\lambda_3$. In the pointlike limit the flow equation \eqref{th6} takes the simple form
\beq \label{ta1}
\partial_{t}\lambda_{3}^{R}=-\frac{4(3+p)}{\sqrt{3}\pi}\left[ \frac{A}{4}+\frac{B}{2}\lambda_{3}^{R}+C(\lambda_{3}^{R})^{2} \right]+2\lambda_{3}^{R},
\eeq
where the rescaled dimensionless coupling $\lambda_3^{R}=\lambda_3k^2$ is introduced. The right-hand side of this differential equation is a quadratic polynomial with constant coefficients. The sign of its discriminant $D$ determines the behavior of the solution of Eq. \eqref{ta1}. For the bosonic problem $D\approx-7.762$ and we obtain a periodic solution of the form $\lambda_3^R\sim \tan[\pi t/T]$ with a period $T=\frac{2\pi}{\sqrt{-D}}$. Note that the renormalization group evolution is governed by a limit cycle scaling. RG limit cycles are the manifestation of the Efimov effect. Indeed, we can give the following  intuitive interpretation of the periodic solution: During the RG flow we encounter three-body atom-dimer bound states, which manifest themselves as divergences of the atom-dimer coupling $\lambda_3^R$. In the unitary limit there are infinitely many of these bound states, which are equidistant in a logarithmic RG energy scale. In this way the continuous scaling symmetry is broken to the discrete scaling subgroup, and the generalized universality emerges.

It is well-known that at the unitarity point, where the scattering length $a$ diverges, the energy spectrum of Efimov trimers forms a geometric series
\beq \label{ta2}
\frac{E_{n+1}}{E_{n}}=\exp(-2\pi/s_{0})
\eeq  
with $E_{n+1}$ and $E_{n}$ denoting neighboring bound state energies. The Efimov parameter $s_{0}$ is given by the solution of a transcendental equation and one finds $s_{0}\approx 1.00624$ \cite{BraatenHammer}. By dimensional arguments we can connect the artificial RG sliding scale $k^{2}$ with the binding energy $E$ as $E\sim k^2$. The proportionality factor disappears in the ratio of the energies and hence the Efimov parameter can be read off from the RG period
\beq \label{ta3}
\frac{k^{2}_{n+1}}{k^{2}_{n}}=\frac{E_{n+1}}{E_{n}}=\exp(-2T) \Rightarrow s_{0}=\frac{\pi}{T}.
\eeq
In the pointlike approximation we obtain $s_{0}\sim 1.393$, which differs from the correct result by $40\%$. Obviously the simple pointlike approximation is too crude to get the correct quantitative agreement. Nevertheless it provides us with the qualitative picture of how the Efimov effect appears as limit cycles in the functional renormalization group framework.

It is remarkable that for $E=0$ the flow equation \eqref{th6} can be solved exactly. Here only the main results will be summarized while for details we refer to \cite{DKS,MFSW}. In order to find the exact solution most easily we perform the redefinition
\beq \label{bf1}
f_{t}(t_{1},t_{2},E)\equiv4(3+p)q_{1}q_{2}\lambda_{3}(q_{1},q_{2},E) \qquad g(t_{1},t_{2})\equiv4(3+p)q_{1}q_{2}G(q_{1},q_{2})
\eeq
and introduce the connected, one-particle reducible atom-dimer vertex
\beq \label{bf3}
\bar{f}_{t}(t_{1},t_{2})=p f_{t}(t_{1},t_{2})+g(t_{1},t_{2}),
\eeq
where from now on we prefer to work with logarithms of momenta $t_{1}=\ln(q_{1}/\Lambda)$ and $t_{2}=\ln(q_{2}/\Lambda)$. The RG scale dependence of the atom-dimer vertex $\bar{f}_{t}(t_1,t_2,E)$ is denoted by the subscript $t$. The renormalization group flow equation and the initial condition for $\bar{f}_t$ are given in matrix notation by
\beq \label{bf5}
\partial_{t}\bar{f}_{t}=-\frac{p}{\sqrt{3}\pi}\bar{f}_{t}\cdot A_{t} \cdot \bar{f}_{t}, \qquad \bar{f}_{t=0}=g,
\eeq
where $A_{t}$ has matrix elements $A_{t}(t_{1},t_{2})=\delta(t-t_{1})\delta(t-t_{2})$ and matrix multiplication denotes integration over $t$. The formal solution of this differential matrix equation reads for $t\in(-\infty,0)$ 
\beq \label{bf7}
\bar{f}_{t}=\left(I+\frac{p}{\sqrt{3}\pi}\int_{0}^{t} ds g\cdot A_{s} \right)^{-1}\cdot g,
\eeq
where $I$ denotes the identity matrix.

In the IR limit $t\to -\infty$, which corresponds to the integration of all quantum fluctuations, $\bar{f}\equiv \bar{f}_{t=-\infty}$ solves the following matrix equation
\beq \label{bf8}
\bar{f}=g+\frac{p}{\sqrt{3}\pi}g\cdot \bar{f}.
\eeq
This is the well-known integral equation for the amputated, connected Green's function obtained by Skorniakov and Ter-Martirosian more than fifty years ago \cite{STM}.

\begin{figure}[t]
\centering
\includegraphics[width=120mm]{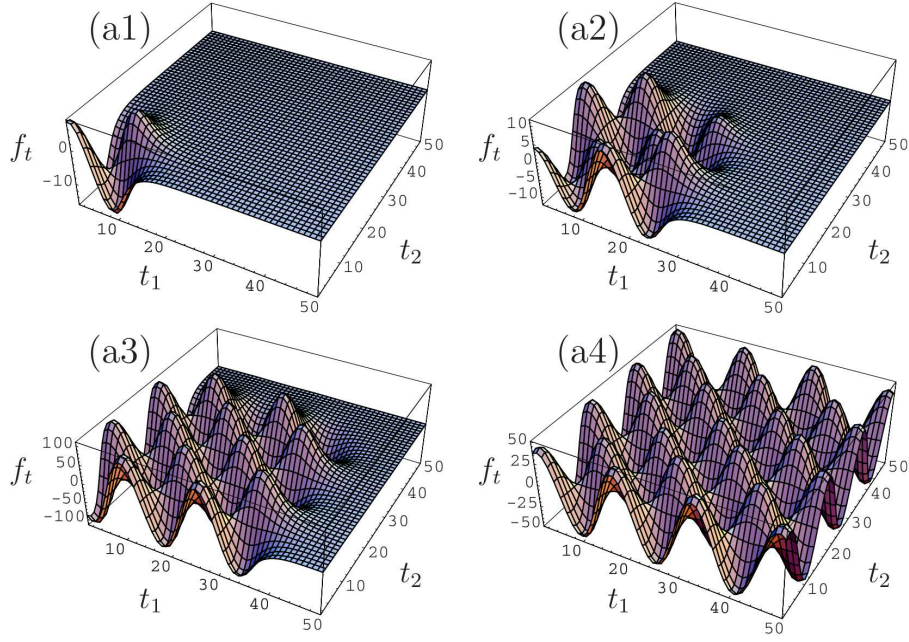}
\caption{\small{The RG evolution of the momentum dependent modified vertex $f_{t}(t_1,t_2)=4(3+p)q_{1}q_{2}\lambda_{3}(q_{1},q_{2},E)$. Spatial momenta $t_1$, $t_2$ and the RG time $t$ are discretized to $N=50$ intervals with a step $\Delta t= 0.4$. The cartoons correspond to the descritized steps $10, 25, 35, 50$.
}}
\label{cartoons}
\end{figure}

\begin{figure}[t]
\centering
\includegraphics[width=70mm]{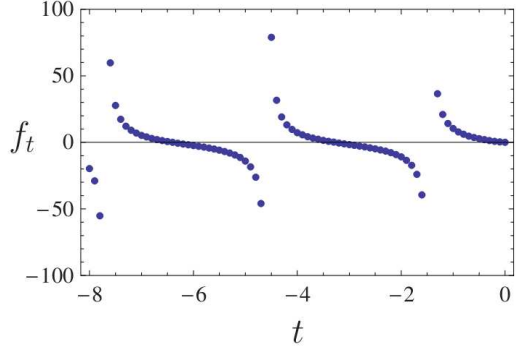}
\caption{\small{Numerical evolution of the point $f_{t}(t_{1}=0,t_{2}=0)$ with the RG time $t$. Renormalization group develops a limit cycle with a period $T_{temp}\approx 3.1$.
}}
\label{RGtimeflow}
\end{figure}
For illustration purposes we discretized Eq. \eqref{bf7} and solved it numerically. A series of cartoons of the RG evolution of the 1PI vertex $f_t(t_1,t_2)$ is depicted in Fig. \ref{cartoons}. In the UV we start with a simple initial condition $f_t(t_1, t_2)=0$ and observe how a periodic structure (with period $T_{\text{spatial}}\approx6.2$) gradually develops. This intricate periodic landscape is completely missed in the pointlike approximation which corresponds to a planar landscape with no $t_1$ and $t_2$ dependence. The RG evolution of the point $f_t(t_1=0, t_2=0)$ of the landscape is depicted in Fig. \ref{RGtimeflow}. Similar to the solution obtained in the pointlike approximation we observe ``temporal`` oscillations. The period of these oscillations $T_{\text{temp}}\approx 3.1$ which leads to the Efimov parameter $s_{0}=\frac{\pi}{T_{\text{temp}}}\approx 1.0$. This is in a good agreement with the ``exact`` result $s_0\approx 1.00624$ obtained from the quantum mechanical treatment and is only limited by our numerical accuracy.

\section{Two-component fermions}
\label{sec:4}
\subsection{Few-body problem}
The solution of the two- and three-body problem, presented in the previous section, can be straightforwardly extended to a system consisting of two fermion species with equal masses $M_{\psi_1}=M_{\psi_2}$ near a broad Feshbach resonance described by the Euclidean action
\begin{equation} \label{tc1}
S = \int_{\tau, \vec x} {\bigg \{ }  \psi_i^*(\partial_\tau - \vec \nabla^2-\mu) \psi_i + \phi^* P_\phi\phi - h (\phi^*\psi_1 \psi_2 + \psi_2^*\psi_1^*\phi)
  {\bigg \}}.
\end{equation}
Here, the two species of elementary fermionic atoms $\psi_{1},\psi_{2}$ are described by Grassmann-valued fields, and $\phi$ is a composite bosonic dimer with the inverse propagator $P_{\phi}$. Wherever equal indices appear summation is implied. Note that in Eq. \eqref{tc1} the dimer $\phi$ is just an auxiliary field which has no own dynamics, i.e. its inverse propagator $P_{\phi}=P_{\phi,k=\Lambda}$ is a constant. This fermionic system has a global $\textrm{SU}(2)\times \textrm{U}(1)$ internal symmetry with $(\psi_{1},\psi_{2})$ transforming as a doublet and $\phi$ as a singlet of the $\textrm{SU}(2)$ flavor subgroup. Two-species fermion systems \eqref{tc1} near a broad Feshbach resonance have for example been realized experimentally using two hyperfine states of $^{6}\text{Li}$ or $^{40}\text{K}$ ultracold atoms.

The two-body problem can be solved analytically following the very same route that we have taken for bosons in subsection \ref{two-bodyproblem}. The only difference for fermions consists in setting $p=-1$ in Eqs. \eqref{eq:twobodydimer1}-\eqref{eq:exactdimerirprop}. Similar to bosons, for $a>0$ there is a universal dimer bound state with the binding energy given by Eq. \eqref{eq:dimerbindingenergy1}.

The solution of the fermionic three-body problem is also directly analogous to the computation for bosons, presented in subsection \ref{three-body problem}. For two-component fermions one arrives at the flow equation \eqref{th4} with coefficients $A=1$, $B=-2$ and $C=1$. The solution of this flow equation is, however, qualitatively very different from the bosonic case. This can be understood already in the pointlike limit, where the flow equation is given by \eqref{ta1}. Crucially, for fermions the discriminant $D$ is positive. We find $D=9.881$, and the solution of the flow equation is of the form $\lambda_3^R\sim \tanh\left(\frac{\sqrt{D}}{2}(t+\kappa) \right)$, where $\kappa$ is fixed by the initial condition. The solution in the pointlike approximation approaches fixed points in the IR and UV. The numerical solution of the full RG flow equation \eqref{th4}, detailed account of which can be found in \cite{MFSW}, supports this finding. In summary, the renormalization group flow does not develop a limit cycle scaling and there is no Efimov effect in the three-body problem for two-component fermions with equal mass. 

\subsection{Many-body physics}
\label{sec:tcmb}
Although the few-body physics of two-component fermions of equal masses is not as rich as for the case of identical bosons, this system exhibits interesting many-body physics \cite{Zwerger,Pitaevskii}. We concentrate on the case of equal densities $\mu=\mu_{\psi_1}=\mu_{\psi_2}$. For small negative scattering length $a\to 0_-$ BCS theory becomes valid. Many-body quantum fluctuations lead to spontaneous $\textrm{U}(1)$ symmetry breaking and a superfluid ground state. Also for small positive scattering length $a\to 0_+$ the ground state will be superfluid, now given by a Bose-Einstein condensate of dimers. In between these two regions one finds a smooth BCS-BEC crossover, including in particular unitarity with $a^{-1}=0$.

The functional renormalization description of this system has been developed in the recent years \cite{DFGPW,BCSBEC,FSW}. It proved to be convenient to work with the Hubbard-Stratonovich transformed microscopic action in Eq.\ \eqref{tc1} and to use a truncation of the flowing action $\Gamma_k$ in terms of a derivative expansion. The most elaborate truncation studied so far \cite{FSW} reads
\begin{eqnarray}
\Gamma_k &= & \int_0^{1/T} d \tau \int d^3 x {\bigg \{ }  \psi_i^* Z_\psi (\partial_\tau-\vec \nabla^2 + m_\psi^2) \psi_i 
 + \phi^* (Z_\phi \partial_\tau - \tfrac{1}{2} A_\phi \vec \nabla^2) \phi \nonumber\\
 &+& U_k(\phi^*\phi) - h (\phi^* \psi_1 \psi_2 + \phi \psi_2^* \psi_1^*) + \lambda_{\phi\psi} \phi^*\phi \psi_i^*\psi_i {\bigg \} }.
\label{tcmb1}
\end{eqnarray}
Here, the coefficients $Z_\psi$, $m_\psi^2$, $Z_\phi$, $A_\phi$, $h$, $\lambda_{\phi\psi}$ and the effective potential $U_k(\phi^*\phi)$ are $k$-dependent quantities. 

The truncation in Eq.\ \eqref{tcmb1} can be used both for the few-body physics (vacuum) and to describe thermal and chemical equilibrium. In the former case one finds the characteristic binding energy of the dimers as in Eq.\ \eqref{eq:dimerbindingenergy1}. Due to simplicity of the truncation it is also rather easy to calculate more complicated observables such as the scattering length of dimers $a_M$. In the limit $a\to 0_+$ the result $a_M/a = 0.59$ \cite{FSW} is in good agreement with the exact result $a_M/a=0.60$ \cite{Petrov} obtained from solving the Schr\"odinger equation. The dimer-dimer scattering was also recently studied with functional renormalization by the Manchester group \cite{Manchester1,Manchester2}. In \cite{Manchester2} all possible local couplings that can be generated in the fermion four-body problem were included and it was found that $a_{M}/a=0.58\pm 0.2$, where the error was estimated from the regulator dependence of the final result.

At non-zero density the flow equations are modified, for example the particle-particle diagram renormalizing the boson propagator in Fig.\ \ref{fig:twobodyflows} has a stronger effect. As a result for small temperatures the effective potential $U_k(\phi^*\phi)$ becomes of a mexican hat form at the macroscopic scale $k=0$. This leads to spontaneous $\textrm{U}(1)$ symmetry breaking and superfluidity. For a more detailed discussion of the flow equations as well as for the results of this approach we refer to ref. \cite{DFGPW,FSW,SFW}.

\section{Three-component fermions with global $\textrm{SU}(3)$ symmetry}
\label{sec:5}
As was discussed in section \ref{sec:4} there is no Efimov effect for two fermion species with equal masses. The main reason for this is Pauli blocking. An obvious question is what happens in the case of three species (flavors) of fermions. The question is of particular interest as systems with three hyperfine states have been realized in recent experimants \cite{Ottenstein08,Huckans09}. In this section we will consider the case where all fermion masses are equal as well as the three pairwise scattering lengths such that the system possesses a global $\textrm{SU}(3)$ internal symmetry. The generalization to the experimentally accessible system of $^6\textrm{Li}$ atoms, where the pairwise scattering lengths are not equal, is discussed in section \ref{sec:6}.

\subsection{Exact solution at unitarity: Vertex expansion}
\label{sec:SU3fermionsvertexexpansion}
For the case of $\textrm{SU}(3)$ symmetric fermions we extend the action \eqref{tc1} to
\begin{equation} \label{su3action}
S = \int_{\tau, \vec x} {\bigg \{ }  \psi_i^*\left(\partial_\tau - \vec \nabla^2-\mu\right) \psi_i+ \phi_i^* \left(\partial_\tau - \vec \nabla^2/2+m_\phi^2\right) \phi_i + \frac{h}{2} \epsilon_{ijk}\left(\phi_i^*\psi_j\psi_k-\phi_i\psi^*_j\psi^*_k\right)    {\bigg \}}.
\end{equation}
Eq.\ \eqref{su3action} is the analog of a two-channel model where the composite bosons $\phi_i$ have a non-trivial propagator already in the microscopic action. The analog of the single channel model in Eq.\ \eqref{tc1} corresponds to the limit of broad resonances $h^2\to \infty$, $m_\phi^2\to \infty$. The inverse propagator of the bosons is then dominated by the constant part $m_\phi^2$. For finite $h$ the action in Eq.\ \eqref{su3action} can describe also Feshbach resonances of finite width.

The three Grassmann valued fields $\psi_i$ for the fermions can be assembled into a vector $\psi=(\psi_1,\psi_2,\psi_3)$ and the bosonic fields $\phi_i$ form a vector $\phi=(\phi_1,\phi_2,\phi_3)\sim(\psi_2\psi_3,\psi_3\psi_1,\psi_1\psi_2)$. The action possesses an $\textrm{SU}(3)\times \textrm{U}(1)$ internal symmetry with $\psi$ transforming as a triplet $\mathbf 3$ and $\phi$ as an anti-triplet $\mathbf{ \bar 3 }$ under the $\textrm{SU}(3)$ flavor subgroup. The three dimers mediate the pairwise scattering in the three possible s-wave scattering channels.

What about Efimov physics in such a system? The answer to this question is in fact quite simple and can be understood with the following quantum mechanical argument: To identify the bound state spectrum one has to solve the three-body Schr\"odinger equation. In the case of fermions the bound state wave function has to be totally antisymmetric. The total wave function can be separated into the orbital and flavor part. If we take the flavor part to be antisymmetric, $(\epsilon_{ijk}|i>|j>|k>)$, the orbital part must be symmetric and therefore has the same symmetry properties as in the bosonic case. As the orbital part solely determines the bound state spectrum in a quantum mechanical calculation, the bound state physics of $\textrm{SU}(3)$ fermions is identical to the bosonic case leading to the same Efimov spectrum. 

A proper analysis of the renormalization group flow equations confirms these expectations. It is more complicated than for identical bosons since two different terms $\sim \phi^*\phi \psi^*\psi$ similar to Eq. \eqref{three1} are allowed by the symmetries. Specifically, the analog to Eqs.\ \eqref{rseq:vebosonstwobody} and \eqref{three1} reads
\begin{eqnarray}\label{fullsu3trunc}
\Gamma_k = \int_{\tau, \vec x} {\bigg \{ }  \psi^\dagger(\partial_\tau - \vec \nabla^2-\mu) \psi+ \phi^\dagger P_\phi \phi+ \frac{h}{2} \epsilon_{ijk}(\phi_i^*\psi_j\psi_k-\phi_i\psi^*_j\psi^*_k) + \lambda_{3a} \phi_i^*\psi_i^*\phi_j\psi_j+\lambda_{3b}\phi^*_i\psi_j^*\phi_i\psi_j   {\bigg \}}.
\end{eqnarray}
We have suppressed here the nonlocal structure (momentum dependence) of the inverse propagator $P_\phi$ and the couplings $\lambda_{3a}$ and $\lambda_{3b}$ for simplicity. 
Nevertheless, in the broad resonance limit $h^2\to \infty$, the solution of the flow equation at $a^{-1}=0$ can be constructed in close analogy to the one presented in Sec.\ \ref{three-body problem}. In particular one finds also a limit cycle with the same Efimov parameter $s_0\approx 1.00624$. For details we refer to \cite{MFSW}.

\subsection{Approximate solution away from unitarity: Derivative expansion}
Although the foregoing discussion formally resolves the question about the Efimov effect for $\textrm{SU}(3)$ fermions, we  want to elaborate a bit more on this case. The solution of the flow equation in terms of the vertex expansion works well (and is numerically exact) at the unitarity point with $a^{-1}=0$ and $E=0$. However, one would like to understand also the region away from this point, especially since for experiments the physics directly at the unitarity point is hard to access. In this subsection we develop a simplified truncation of the flowing action in terms of a derivative expansion. 

Based on the ansatz \eqref{fullsu3trunc} our first goal is to find a `minimal' truncation which still captures the essential features of Efimov physics. Due to fermion statistics the Efimov trimer is expected to be a $\textrm{SU}(3)$ singlet. Only the term proportional to $\lambda_{3a}$ in Eq. \eqref{fullsu3trunc} can have a bound state pole due to the combination of all three flavours. From this perspective, the term proportional to $\lambda_{3b}$, which plays a similar role as in the two component fermion case, is expected to be less relevant for the description of the three-body bound state. We expect it only to affect quantitative results but not the qualitative features and we will neglect it from here on. 

Instead of keeping the full momentum and energy dependence of $P_\phi$ and $\lambda_{3a}$ as in the vertex expansion \eqref{fullsu3trunc} we will employ a derivative expansion
\begin{eqnarray}\label{su3trunc}
\Gamma_k &=&\int_{\tau, \vec x} {\bigg \{ }  \psi^*_i(\partial_\tau - \vec \nabla^2-\mu) \psi_i+ \phi_i^*\left[A_\phi\left( \partial_\tau - \vec \nabla^2/2 \right)+m_\phi^2 \right] \phi_i + \chi^*\left[A_\chi\left( \partial_\tau - \vec \nabla^2/3 \right)+m_\chi^2 \right] \chi \nonumber\\
&&\quad\quad+ \frac{h}{2} \epsilon_{ijk}\left(\phi_i^*\psi_j\psi_k-\phi_i\psi^*_j\psi^*_k\right) +g\left(\phi_i^*\psi_i^*\chi-\phi_i\psi_i\chi^*\right)
{\bigg \}}.
\end{eqnarray}
The inverse dimer propagator is now approximated by a gradient expansion in leading order in momentum and energy with a gap term $m_\phi^2$ and wave function renormalization $A_\phi$. The truncation \eqref{su3trunc} furthermore contains a single component fermionic field $\chi$ which is a singlet under $\textrm{SU}(3)$ and represents the totally antisymmetric combination $\psi_1\psi_2\psi_3$. The Efimov trimer field $\chi$, which we also call sometimes the `trion', mediates the original atom dimer interaction $\sim \lambda_{3a}$ and is introduced by a Hubbard-Stratonovich transformation analogously to the dimer field introduced in section \ref{truncations}. Similar to the dimer we approximate the inverse propagator for the field $\chi$ in terms of a derivative expansion. 

The Yukawa-type term proportional to $g$ describes how the composite field $\chi$ is formed from its constituents.  At the microscopic scale $k=\Lambda$ we choose $g=0$, $A_\chi=1$ and $m_\chi^2\to\infty$. The field $\chi$ is then an auxiliary field that decouples from the other fields. With this choice we recover the original microscopic action \eqref{su3action} in the broad resonance limit. In the next subsection we will relate the initial values of the dimer gap $m_\phi^2(\Lambda)$ and the Yukawa coupling $h$ to the scattering properties of the elementary fermions $\psi_i$.

\subsubsection{Two-body problem}
We proceed as in section \ref{sec:2} by subsequently solving the flow equations in the two- and three-body sector. In the following analysis we will employ the regulator $R_k=r(k^2-\textbf p^2)\theta(k^2-\textbf p^2)$ with $r=1$ for the fermions $\psi_i$, $r=A_\phi/2$ for the bosons $\phi_i$ and $r=A_\chi/3$ for the trion field $\chi$. From the arguments presented in Sect.\ \ref{two-bodyproblem} it follows that the fermion propagator and the Yukawa coupling $h$ are not renormalized. The flow equation for the boson gap parameter reads
\begin{equation}\label{mphifeq}
\partial_t  m_\phi^2 = \frac{ h^2}{6\pi^2}  \frac{k^5}{(k^2-\mu)^2},
\end{equation}
and for the boson wave function renormalization we obtain
\begin{equation}\label{aphifeq}
\partial_t A_\phi = -\frac{h^2}{6\pi^2}\frac{k^5}{(k^2-\mu)^3}.
\end{equation}
Eqs. \eqref{mphifeq}, \eqref{aphifeq} can be solved by direct integration, for details we refer to ref.\ \cite{FSMW}.

The initial values for $A_\phi$, $m_\phi^2$, and $h$ can be related to the scattering properties of the elementary fermions close to a Feshbach resonance. We can use the freedom to perform a rescaling of the field $\phi_i$ and choose $A_\phi(\Lambda)=1$. To fix the remaining parameters we note that the scattering length between the fermions is given by $a=-\frac{h^2}{8\pi m_\phi^2}$ where the couplings are evaluated at the macroscopic (IR) scale $k=0$ and for vanishing $\mu$. This fixes the initial value of $m_\phi^2$ to be
\begin{equation}
m_\phi^2(\Lambda)=-\frac{h^2}{8\pi}a^{-1}+\frac{h^2}{6\pi^2}\Lambda-2\mu,
\end{equation}
where the term $-2\mu$ accounts for the fact that the dimer consists of two fermions.

The value of the Yukawa coupling $h$ is related to the width of the Feshbach resonance. The scattering length close to a Feshbach resonance at magnetic field $B_0$ and width $\Delta B$ can be parametrized by \cite{Zwerger,Kokkelmans}
\begin{equation}
a=\frac{a_{bg} \Delta B}{B-B_0}
\end{equation}
with $a_{bg}$ the background scattering length. Furthermore, we may identify the boson gap $m_\phi^2(k=0,\mu=0)$ with the energy of the closed channel molecule, $E_c= \mu_M(B-B_0)$, with $\mu_M$ being its magnetic moment (for more details see \cite{DFGPW}). From this one can see that the Yukawa coupling $h^2$ is proportional to the width of the Feshbach resonance:
\begin{equation}
h^2=-8\pi a_{bg}\mu_M \Delta B.
\end{equation} 
Broad Feshbach resonances correspond to $h^2\to \infty$. However, we can study also the crossover from narrow to broad Feshbach resonances within the two-channel model Eq.\ \eqref{su3action}.

\subsubsection{Three-body problem and energy spectrum}\label{sec:tbp}
The flow equations of the three-body sector read
\begin{equation}
\partial_t m_\chi^2=\frac{6g^2}{\pi^2}\frac{A_\phi k^5}{(3A_\phi k^2-2A_\phi \mu +2 m_\phi^2)^2},
\label{eq:flowmchi}
\end{equation}
and
\begin{equation}
\partial_t g = m_\chi^2\partial_t \alpha -\frac{2g h^2}{3\pi^2}\frac{k^5}{(k^2-\mu)^2}\frac{(6A_\phi k^2-5A_\phi \mu +2 m_\phi^2)}{(3A_\phi k^2-2A_\phi \mu +2 m_\phi^2)^2}.
\label{eq:flowg2}
\end{equation}
\begin{figure}[t!]
\begin{minipage}[c]{0.5\textwidth}
\centering
\includegraphics[width=0.9\textwidth]{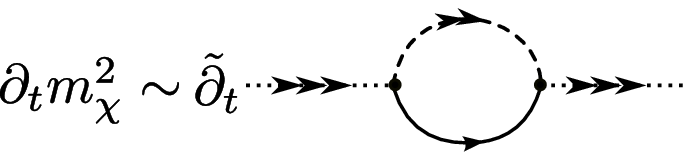}
\end{minipage}%
\begin{minipage}[c]{0.5\textwidth}
\centering 
\includegraphics[width=0.95\textwidth]{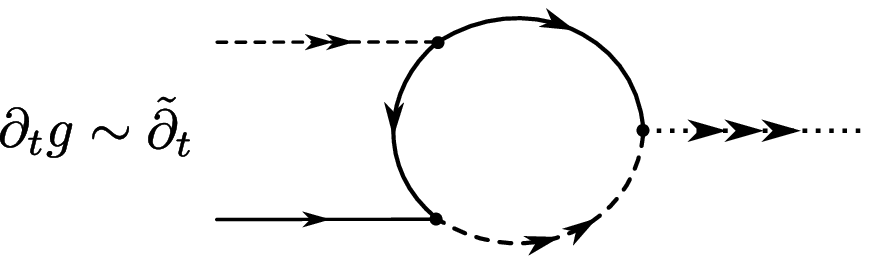}
\end{minipage}
\caption{The beta functions of the three-body couplings $m_\chi^2$ and $g$ depicted as Feynman diagrams. The dotted lines with three arrows denote the composite trimer field $\chi$.}
\label{fig:gandmchiflow}
\end{figure}
While the expression in Eq. \eqref{eq:flowmchi} and the second term in Eq. \eqref{eq:flowg2} are straightforwardly found from the diagrams shown in Fig. \ref{fig:gandmchiflow}, the first term in Eq. \eqref{eq:flowg2} needs further clarification. 

\begin{figure}[t!]
\centering
\includegraphics[width=0.4\textwidth]{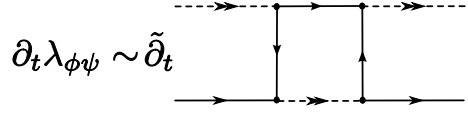}
\caption{The box diagram leads to the regeneration of the atom-dimer vertex $\lambda_{\phi\psi}$ in the RG flow. Using a $k$-dependent Hubbard-Stratonovich transformation it is absorbed in the flow of the Yukawa coupling $g$.}
\label{fig:boxdiagram}
\end{figure}

Although the original atom-dimer interaction $\sim \lambda_{3a}$ has been replaced on the UV scale by the trion exchange using the Hubbard-Stratonovich transformation, an atom-dimer interaction term of the form 
\begin{equation}
\int_{\tau, \vec x} \lambda_{\phi\psi} \phi^*_i \psi_i^* \phi_j \psi_j
\end{equation}
builds up again in the RG flow due to the box diagram shown in Fig. \ref{fig:boxdiagram}. As it turns out, it is essential to include this evolution in our RG treatment to recover Efimov physics. In order to do so, we use a scale dependent Hubbard-Stratonovich transformation, a method developed in \cite{Bosonization,FWCompositeOperators}. By this we absorb the regenerated atom-dimer coupling $\lambda_{\phi\psi}$ into the flow of the Yukawa coupling $g$. 

As was shown in \cite{Bosonization,FWCompositeOperators}, the Wetterich equation gets supplemented by an additional term from the scale dependence of the Hubbard-Stratonovich transformation. We choose this scale dependence such that the box diagram in Fig. \ref{fig:boxdiagram} gets cancelled. This leads to
\begin{equation}
\partial_t \alpha = -\frac{h^4}{12\pi^2 g}\frac{k^5}{(k^2-\mu)^3}\frac{(9A_\phi k^2-7A_\phi \mu +4 m_\phi^2)}{(3A_\phi k^2-2A_\phi \mu +2 m_\phi^2)^2}.
\label{eq:alpha}
\end{equation}
This determines the first term in Eq. \eqref{eq:flowg2} and with this choice $\lambda_{\phi\psi}\equiv 0$ is enforced on all scales $k$. The scattering between fermions and bosons is now described exclusively by the exchange of the (scale dependent) trion bound state $\chi$.

We now investigate the energies of the different states -- elementary fermions $\psi_i$ (``atoms''), composite bosons $\phi_i$ (``dimers'') or composite fermions $\chi$ (``trimers'') -- as a function of the model parameters (e.g. the fermion scattering length $a$). In general it is not possible to find a simple analytical solution to the flow equations \eqref{eq:flowmchi}, \eqref{eq:flowg2} and we have to resort on a numerical solution. To find the energy of the lowest excitation we adjust the chemical potential $\mu\leq0$ such that the lowest excitation of the vacuum is gapless at $k=0$. 

For a finite UV cutoff scale $\Lambda$ we find that for $a^{-1}\to -\infty$, the lowest energy state is given by the atoms which implies $\mu=0$, $m_\phi^2>0$, and $m^2_\chi>0$. No bound states are present in this regime.

In contrast, for $a^{-1}\to \infty$, the dimer $\phi$ is the lowest energy state, $\mu<0$, $m_\phi^2=0$, and $m^2_\chi>0$.  In fact, we can use our analytical solution of the two-body sector to obtain the dimer energy. In the limit $\Lambda/|\mu|\to\infty$ we find
\begin{equation}
\label{eq:bindingenergydimer}
\mu =-\left(\frac{ h^2}{32 \pi}-\sqrt{\frac{h^4}{(32 \pi)^2}+\frac{a^{-1} h^2}{16 \pi}}\right)^2.
\end{equation}
which gives the binding energy of the weakly bound dimer state $E_D=2 \mu$. For finite $h$ and the choice $A_\phi(\Lambda)=1$ the dimer binding energy also depends on the ``effective range'' 
\begin{equation}
r_\textrm{eff}=-\frac{32\pi}{h^2},
\end{equation}
according to
\begin{equation}
E_D=2\mu=-\frac{2}{r_\textrm{eff}}\left(1-\sqrt{1-\frac{2 r_\textrm{eff}}{a}}\right)^2.
\end{equation}
In the broad resonance limit (vanishing effective range) one recovers the well-known result $E_D=-2/a^2=-\hbar^2/(M_\psi a^2)$.

\begin{figure}[t!]
\begin{minipage}[c]{0.5\textwidth}
\centering
\includegraphics[width=0.9\textwidth]{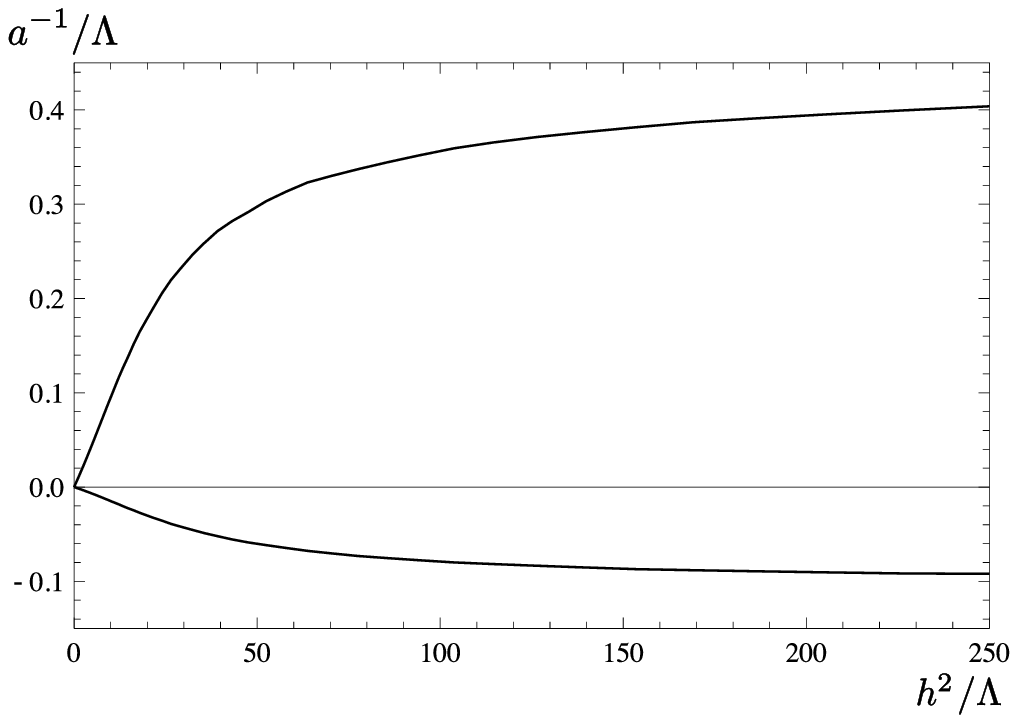}
\end{minipage}%
\begin{minipage}[c]{0.5\textwidth}
\centering 
\includegraphics[width=0.95\textwidth]{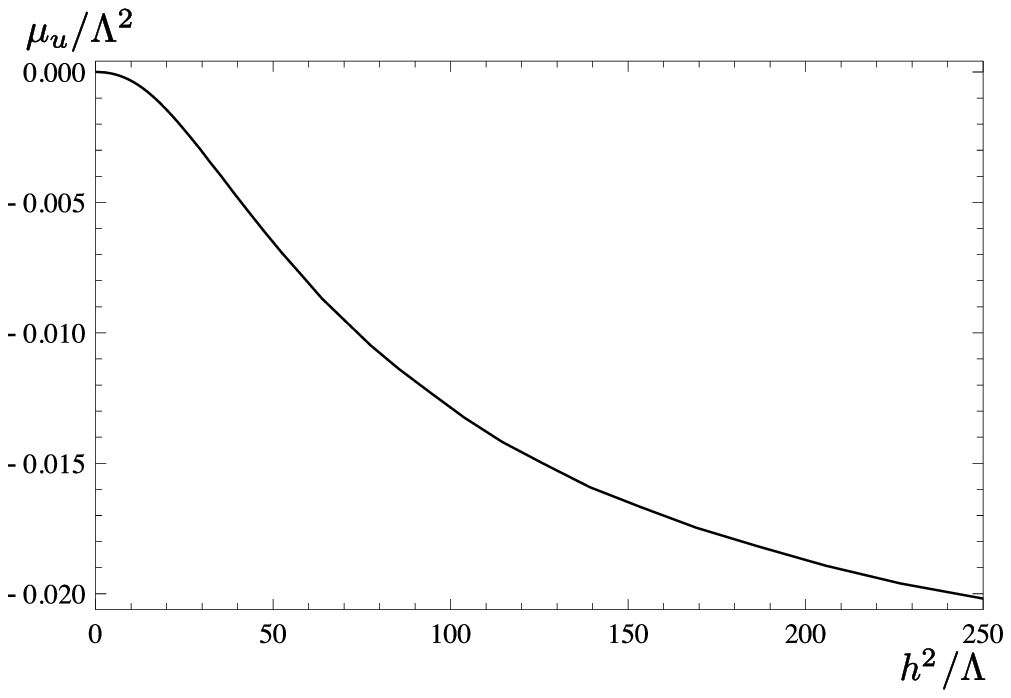}
\end{minipage}
\caption{Crossover from narrow to broad Feshbach resonances. left: Scaling of the critical scattering length $a^{-1}_{c1}<0$ ($a^{-1}_{c2}/4>0$) at which the lowest Efimov trimer becomes degenerate with the atom (atom-dimer) threshold in dependence on the width of the Feshbach resonance given by $h$. right: Analogous scaling of the lowest trimer binding energy given by the chemical potential $\mu_u$ at unitarity.}
\label{fig:scaling}
\end{figure}

Let us investigate to the regime of strong interaction close to the point with $a^{-1}=0$. In a region $a^{-1}_{c1}\leq a^{-1}\leq a^{-1}_{c2}$, we find that the lowest energy is given by the trion, such that $\mu<0$, $m_\phi^2>0$, and $m^2_\chi=0$ for $k=0$. In this region the lowest energy state is a bound state of three fundamental fermions $\chi \sim \psi_1 \psi_2 \psi_3$. The critical scattering length $a^{-1}_{c1}$ ($a^{-1}_{c2}$), at which the Efimov trimer becomes degenerate with the atom (atom-dimer) threshold, depends on the effective range. For narrow resonances $h^2/\Lambda\ll1$ we find a linear increase with $h^2$ as  $a^{-1}_{c1}=-0.0013h^2$ and  $a^{-1}_{c2}= 0.0085 h^2$. For broad Feshbach resonances, $h^2/\Lambda\gg1$, the depth of the trimer state becomes independent of the effective range and is determined solely by the UV cutoff scale $a^{-1}_{c1}$, $a^{-1}_{c2}\sim\Lambda$ such that the cutoff scale can be viewed as the additional three-body parameter needed to determine the exact position of the universal Efimov trimer state \cite{BraatenHammer}. This crossover behavior from narrow to broad resonances is shown in Fig. \ref{fig:scaling} (left).

In Fig. \ref{fig:scaling} (right) we show our result for the energy per atom $\mu_u$ of the trimer as a function of $h^2$ at unitarity $a^{-1}=0$. For narrow resonances we find $\mu_u=-2.5 \times 10^{-6}h^4$. In the broad resonance limit $h^2\to \infty$ also $\mu_u$ depends only on the three-body parameter given by the cutoff scale, $\mu_u\sim\Lambda^2$.

From the work of Efimov \cite{Efimov70}, and as we seen in section \ref{sec:SU3fermionsvertexexpansion}, one expects a whole tower of infinitely many Efimov trimer states close to the resonance. In the following we will calculate this excitation spectrum. In order to do so, we make the following observation. First consider the inverse propagators of the atoms, dimers and trimer after  analytical continuation, $\tau=i t$, to real frequencies
\begin{eqnarray}
P_\psi&=&-\omega_\psi+\vec p^2-\mu,\notag\\
P_\phi/A_\phi&=&-\omega_\phi+\frac{\vec{p}^2}{2}-2\mu+\nu_\phi(\mu+\omega_\phi/2-\vec{p}^2/4), \notag\\
P_\chi/A_\chi &=&-\omega_\chi+\frac{\vec{p}^2}{3}-3\mu+\nu_\chi(\mu+\omega_\chi/3-\vec{p}^{2}/9).\label{eq:propagators}
\end{eqnarray}
We note that $\nu_\phi$ and $\nu_\chi$ can dependent only on the particular combinations of $\mu$, $\omega$, and $\textbf p^2$ indicated above. This is a consequence of an additional symmetry of the real-time microscopic action $S=\Gamma_{\Lambda}$ under the time-dependent $\textrm{U}(1)$ transformation $\psi\to e^{iEt}\psi$, $\phi\to e^{i2Et}\phi$, and $\chi\to e^{i3Et}\chi$ if we also simultaneously change the chemical potential $\mu\to\mu+E$. In vacuum we do not expect any anomalies, such that also the quantum effective action $\Gamma=\Gamma_{k=0}$ possesses this symmetry. As consequence any dependence on $\mu$ is accompanied by a corresponding frequency dependence and vice versa. This allows only the particular combinations $\omega_\psi+\mu$, $\omega_\phi+2\mu$, and $\omega_\chi+3\mu$ to appear. The additional relation between frequency and momenta is a consequence of Galilean invariance.

The dispersion relations are obtained from the poles of the propagators, that is from solving $P_\phi(\omega_\phi=\mathcal{E}_\phi)=0$ and analogous for the atoms and the trion. The dimer dispersion for example reads
\begin{equation}
\mathcal{E}_\phi(\textbf p^2)=\frac{\textbf p^2}{2}-2\mu+\nu_\phi(\mathcal{E}_\phi/2+\mu-\textbf p^2/4)=\frac{\textbf p^2}{2}+\frac{m_\phi}{A_\phi}.
\end{equation}
To obtain the dimer and trimer energy levels $E_\phi$ and $E_\chi$ we have to substract two or three times the atom energy $E_\psi=-\mu$ from the dispersion relation at rest such that 
\begin{eqnarray}
E_\phi&=&\mathcal{E}_\phi(\textbf p^2=0)+2 \mu=\nu_\phi(\omega_\phi=\mathcal E_\phi,\textbf p^2 =0,\mu),\notag\\
E_\chi&=&\mathcal{E}_\chi(\textbf p^2=0)+3\mu=\nu_\chi(\omega_\chi=\mathcal E_\chi,\textbf p^2 =0,\mu).
\label{eq:energylevels}
\end{eqnarray}
Here $\mu$ is the effective chemical potential which accounts for an shift in the atom energy using the time-dependent $\textrm{U}(1)$ symmetry. The ground and excited states are obtained by the condition that the particular particles are gapless in the IR which corresponds to a gapless dispersion, i. e. for example $\mathcal{E}_\phi(\textbf p^2=0)=0$. This means that the corresponding gap parameter has to vanish in the infrared, i. e. $\mu=0$, or $m_\phi^2=0$, or $m^2_\chi=0$. In practice, we extract the bound state energy levels $E_\phi=2\mu$, $E_\chi=3\mu$ by calculating the gaps $m_\phi^2$ and $m_\chi^2$ for $k=0$ such that they vanish for a specific choice of $E_\psi=-\mu$. 

\begin{figure}[t!]
\centering
\includegraphics[width=0.6\textwidth]{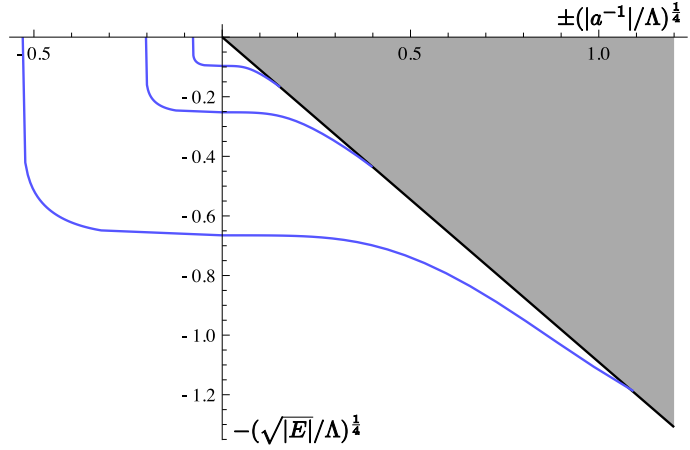}
\caption{The Efimov spectrum for $\textrm{SU}(3)$ fermions. We plot the lowest three Efimov trimer levels $E_\chi$ (blue). The dimer binding energy is shown as black curve. To improve the visibility of the energy levels, we rescale both the dimensionless inverse scattering length $a^{-1}/\Lambda$ and the dimensionless energy $E/\Lambda^2$. We have chosen the UV cutoff $\Lambda=a_0$ and the Yukawa coupling $h^2\Lambda=100$ as appropriate for $^6\textrm{Li}$ atoms in the $m_F=1/2$, $m_F=-1/2$ channel close to the Feshbach resonance at $B_0\sim 810\, \textrm{G}$.}
\label{fig:spectrum}
\end{figure}

As it turns out for the trimer there is not only the ground state solution which we discussed before but a whole infinite set of possible solutions corresponding to the infinite tower of universal Efimov trimer states. The numerical result for the dimer energy level $E_\phi$ and the lowest three Efimov trimer levels $E_\chi$ as function of $a^{-1}$ measured in units of $\Lambda$ is shown in Fig. \ref{fig:spectrum}. In this figure we used a convenient rescaling to make the excited Efimov states visible. We have identified the UV cutoff $\Lambda$ with the inverse Bohr radius, $\Lambda=a_0^{-1}$, and have additionally chosen $h^2=100\Lambda$. This choice corresponds to the width of the Feshbach resonance of $^6\textrm{Li}$ atoms in the $m_F=1/2$, $m_F=-1/2$ channel relevant for experiments \cite{Bartenstein}.

\subsubsection{Limit cycle scaling at the unitarity point}
Already in section \ref{sec:2} we have discussed the especially interesting unitarity point in the spectrum where the scattering length diverges and the atom energy is at the threshold, $a^{-1}=E_\psi=-\mu=0$. It is instructive to consider this limit also within the derivative expansion as it allows us to obtain analytical results. At the unitarity point all length scales drop out of the problem and one obtains a simple scaling solution for the flow equations. In this limit the flow equations for the three-body sector simplify and read
\begin{eqnarray}
\nonumber
\partial_t m_\chi^2 &=& \frac{36}{25}  g^2\frac{k^2}{ h^2}\\
\partial_t g^2 &=& -\frac{64}{25} g^2-\frac{13}{25}m_\chi^2 \frac{h^2}{k^2}.
\label{eq:mgsystem}
\end{eqnarray}
where we used $A_\phi(k)=h^2/(6\pi^2k)$ and $m_\phi^2=A_\phi k^2$ for $k\to 0$.
Using $g(\Lambda)=0$ as initial coupling we find the oscillatory, analytical solution
\begin{eqnarray}
m_\chi^2(k)&=&\left(\frac{k}{\Lambda}\right)^{-\frac{7}{25}} m_\chi^2(\Lambda)\Big[\cos\left(s_0 \ln\frac{k}{\Lambda}\right)+\frac{7}{\sqrt{419}}\sin\left( s_0 \ln \frac{k}{\Lambda}\right)\Big],\notag \\
g^2(k)&=&-\frac{13 \,h^2}{\sqrt{419}\,k^2}\left(\frac{k}{\Lambda}\right)^{-\frac{7}{25}} m_\chi^2(\Lambda) \sin\left(s_0 \ln \frac{k}{\Lambda} \right),
\label{eq:mgsolution}
\end{eqnarray}
with  $s_0=\sqrt{419}/25\approx0.82$.\footnote{Using the analogous derivative expansion for bosons, where there exists only a single term describing the atom-dimer interaction $\sim\phi^*\psi^*\phi\psi$, one finds $s_0\approx0.93$ \cite{RM}.} From this solution we see that $m_\chi^2(\Lambda)$ drops out of the ratio $g^2/m_\chi^2$ which determines the atom-dimer interaction $\lambda_{3a}\sim g^2/m^2_\chi$. The flow of $m_\chi^2$ and $g^2$ describes a closed limit cycle with the frequency given by $s_0$. Considering the simplicity of our approximation this is in reasonable agreement with the `exact' result, $s_0\approx1.00624$. The limit cycle goes on `forever' only exactly at the unitarity point and at the threshold energy $\mu=0$. A finite $a^{-1}$ as well as a finite atom energy $E_\psi=-\mu$ act as infrared cutoffs, which stop the limit cycle flow once $k\approx |\mu|$ or $k\approx 3\pi |a^{-1}|/4$, respectively.

\subsection{Many-body physics}
The simple truncation of the exact functional renormalization group equation in terms of a derivative expansion yields a detailed picture of the spectrum of bound states and scattering in the system with three species of fermions with $\textrm{SU(3)}$ symmetry. Depending on the scattering length $a$, the lowest energy state may be a fundamental fermion (atom) or a bosonic molecule (dimer) or a composite fermion formed from three atoms (trion). Trions are stable (lowest energy state) in the interval $(a^{(1)}_{c1})^{-1}<a^{-1}<(a^{(1)}_{c2})^{-1}$.

We have so far only considered the flow equations in vacuum, but we can already infer some interesting features of what happens at non-zero density for the many-body ground state at $T=0$. The qualitative properties for small density follow from our computations and simple arguments of continuity. Importantly, we find that a novel ``trion phase'' separates the BEC phase and the BCS phase  \cite{FSMW}.

For nonvanishing density, the chemical potential $\mu$ increases above its vacuum value $\mu_0$. At $T=0$ for $a^{-1}>(a_{c2})^{-1}$ the BEC phase occurs for small density. Due to the symmetries of the microscopic action, the effective potential $U(\rho)$ for the bosonic field depends only on the $\textrm{SU(3)}\times\textrm{U(1)}$ invariant combination $\rho=A_\phi \phi^\dagger\phi= A_\phi (\phi_1^*\phi_1+\phi_2^*\phi_2+\phi_3^*\phi_3)$. It reads for small $\mu-\mu_0$
\begin{equation}
U(\rho)=\frac{\lambda_\phi}{2}\rho^2-2(\mu-\mu_0)\rho,
\label{eq::effpot}
\end{equation}
where we use the fact that the term linear in $\rho$ vanishes for $\mu=\mu_0$, i.e. $m_\phi^2(\mu_0)=0$. The minimum of the potential shows a non-zero condensate
\begin{equation}
\rho_0=\frac{2(\mu-\mu_0)}{\lambda_\phi},\qquad  U(\rho_0)=-\frac{2}{\lambda_\phi}(\mu-\mu_0)^2,
\end{equation}
with density
\begin{equation}
n=-\frac{\partial U(\rho_0)}{\partial \mu}=\frac{4}{\lambda_\phi}(\mu-\mu_0).
\end{equation}

For the BCS phase for $a<a_{c1}<0$, one has $\mu_0=0$. Non-zero density corresponds to positive $\mu$, and in this region the renormalization flow drives $m_\phi^2$ always to zero at some finite $k_c$, with BCS spontaneous symmetry breaking ($\rho_0>0$) induced by the flow for $k<k_c$. Both the BEC and BCS phases are therefore characterized by superfluidity with a non-zero $\rho_0=A_\phi \phi^{\dagger}\phi$.

As an additional feature compared to the BCS-BEC crossover for a Fermi gas with two components, the expectation value for the bosonic field $\phi$ in the three component case also breaks the ``spin`` $\text{SU(3)}$ symmetry. Due to the analogy with QCD this is called ``Color Superfluidity'' \cite{Honerkamp}, see also \cite{Wilczek}. For any particular direction of the expection value  of $\phi$ a continuous symmetry $\text{SU(2)}\times\text{U(1)}$ remains. According to the symmetry breaking $\text{SU(3)}\times\text{U(1)}\to \text{SU(2)}\times\text{U(1)}$, the effective potential has five flat directions.

For two component fermions the BEC and BCS phases are not separated, since in the vacuum either $\mu_0=0$ or $m_\phi^2=0$. There is no phase transition, but rather a continuous crossover. For three component fermions, however, we find a new trion phase for $(a^{(1)}_{c1})^{-1}<a^{-1}<(a^{(1)}_{c2})^{-1}$. In this region the vacuum has $\mu_0<0$ and $m_\phi^2>0$. The atom fluctuations are cut off by the negative chemical potential and do not drive $m_\phi^2$ to zero, such that for small density $m_\phi^2$ remains positive. Adding a term $m_\phi^2 \rho$ to the effective potential \eqref{eq::effpot} we see that the minimum remains at $\rho_0=0$ as long as $m_\phi^2>2(\mu-\mu_0)$. No condensate of dimers occurs in the trion phase. The BEC and BCS phases, that show both extended superfluidity through a spontaneous breaking of the $\text{SU(3)}\times\text{U(1)}$ symmetry, are now separated by a phase where $\rho_0=0$, such that the $\text{SU(3)}\times\text{U(1)}$ symmetry remains unbroken (or as we argue below will be only partially broken).

Deep in the trion phase, e.g. for very small $|a^{-1}|$, the atoms and dimers can be neglected at low density since they both have a gap. The thermodynamics at low density and temperature is determined by a single species of fermions, the trions. In our approximation it is simply given by a non-interacting Fermi gas, with fermion mass $3M$ and chemical potential $3(\mu-\mu_0)$. Beyond our approximation, we expect that trion interactions are induced by fluctuations. While local trion interactions $\sim (\chi^*\chi)^2$ are forbidden by Fermi statistics, momentum dependent interactions are allowed. These may, however, be ``irrelevant interactions'' at low density, since also the relevant momenta are small such that momentum dependent interactions will be suppressed. Even if attractive interactions would induce a trion-trion condensate, this has atom number six and would therefore leave a $\textrm{Z}_6$ subgroup of the $\textrm{U}(1)$ transformations unbroken, in contrast to the BEC and BCS phase where only $\textrm{Z}_2$ remains. Furthermore, the trions are $\text{SU(3)}$-singlets such that the $\text{SU(3)}$ symmetry remains unbroken in the trion phase. The different symmetry properties between the possible condensates guarantee true quantum phase transitions in the vicinity of $a_{c1}$ and $a_{c2}$ for small density and at $T=0$. We expect that this phase transition also extends to small non-zero temperature.

While deep in the trion phase the only relevant scales are given by the density and temperature, and possibly the trion interaction, the situation gets more complex close to quantum phase transition points. For $a\approx a_{c1}$ we have to deal with a system of trions and atoms, while for $a\approx a_{c2}$ a system of trions and dimers becomes relevant. The physics of these phase transitions is not clear now, it might be complex and interesting.

\section{Three-component fermions without $\textrm{SU}(3)$ symmetry}\label{sec:6}
\begin{figure}[t!]
\centering
\includegraphics[width=0.55\textwidth]{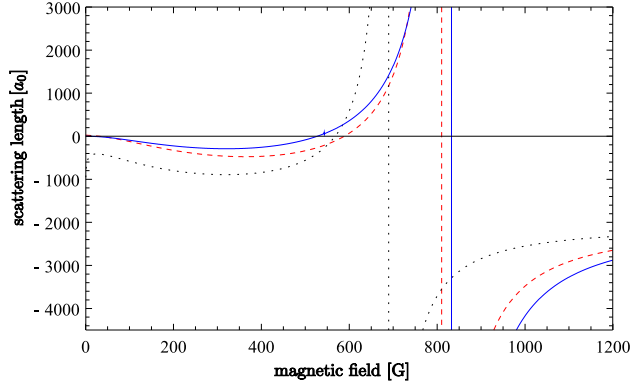}
\caption{Two-body scattering lengths $a_{12}$ (solid), $a_{23}$ (dashed), and $a_{13}$ (dotted) in dependence of the magnetic field $B$ of $^6\textrm{Li}$ atoms in the lowest three hyperfine states. These curves were calculated by P.~S.~Julienne \cite{Bartenstein} and are taken from Ref. \cite{Ottenstein08}. }
\label{fig:scattlength}
\end{figure}

Quite recently three-component Fermi gases have been realized with ultracold atoms \cite{Ottenstein08,Huckans09}. In these experiments $^6\textrm{Li}$ atoms are prepared in three hyperfine states. Using external magnetic fields the scattering length of the different pair of atoms $(1,2)$, $(1,3)$, and $(2,3)$ can be tuned due to the presence of three Feshbach resonances. In the $\textrm{SU}(3)$ invariant system discussed before all pairs of atoms had equal scattering lengths. This is not valid in the case of $^6\textrm{Li}$ atoms as can be seen from the scattering lengths profiles shown in Fig. \ref{fig:scattlength}. Although the Feshbach resonances are close to each other, they are still at different magnetic fields and also the profiles of the scattering lengths differ for the different pairs of atoms. In order to compare our calculations with the experimental observations we have therefore to extend the model discussed in section \ref{sec:5}.

The experiments by Ottenstein \textit{et al.} \cite{Ottenstein08} and Huckans \textit{et al.} \cite{Huckans09} concentrated on the magnetic field region between $B=100\, \textrm{G}$ and $B=550\, \textrm{G}$ where all scattering lengths are rather large and negative. The main observation was a distinct feature in the atom loss rate which could be attributed to a three-body process and shows up in the three-body coefficient $K_3$ which is defined by the loss rate:
\begin{equation}
\frac{dn_i}{dt}=K_3 n_i^3,
\end{equation}
where $n_i$ denotes the density of the various atom species. The measured data points from \cite{Ottenstein08} are shown in Fig. \ref{fig:K3} (right) together with our numerical results which will be derived below.

Already from the work of Efimov \cite{Efimov70} it is known that trimer bound states should persist in systems were the pairwise scattering lengths of the three involved particles differ from each other as long as at least to of them are large in magnitude. This is the case in the regime investigated by the experiments and in the following we will show how the observed loss features can be explained by Efimov physics using functional renormalization \cite{FScW}. Similar predictions have been made using different methods by Braaten \textit{et al.} \cite{Braatenli6} and Naidon und Ueda \cite{Naidon}.

\subsection{Derivative expansion}
First consider the model which we will use to describe the system. We generalize the truncation given by Eq. \eqref{su3trunc} presented in section \ref{sec:5}. The average action then reads
\begin{eqnarray}
\nonumber
\Gamma_k &=& \int_x {\bigg \{} \sum_{i=1}^3 \psi_i^*(\partial_\tau-\mathbf\nabla^2-\mu)\psi_i+\sum_{i=1}^3 \phi_i^*\left[A_{\phi i}\left(\partial_\tau-\vec\nabla^2/2\right)+m_{\phi i}^2\right]\phi_i+ \chi^*\left[A_\chi \left(\partial_\tau-\vec\nabla^2/3\right)+m_\chi^2\right]\chi\\
&& +\sum_{i,j,k=1}^3\frac{h_i}{2} \epsilon_{ijk}\left( \phi_i^* \psi_j\psi_k-\phi_i\psi_j^*\psi_k^*\right)+ \sum_{i=1}^3 g_i\left(\phi_i^* \psi_i^* \chi-\phi_i \psi_i \chi^*\right) {\bigg \}},
\label{nosu3trunc}
\end{eqnarray}
The generalization consists in allowing for three different inverse boson propagators determined by the six couplings $A_{\phi i}$ and $m^2_{\phi i}$, $i=1,2,3$, and the three independent Yukawa couplings $h_i$. This enables us to describe the distinct scattering properties of the three channels. Furthermore we allow for three couplings $g_i$ which couple a dimer and a fermion to the trimer bound state $\chi$. As in section \ref{sec:5}, the trimer field $\chi$ mediates the effective interaction between dimers and atoms $\lambda_{ij}^{(3)}$ such that for vanishing center of mass momentum one has
\begin{equation}
\lambda_{ij}^{(3)}=-\frac{g_i g_j}{m_\chi^2}.
\end{equation}
By permitting three independent couplings $g_i$ we are able to account for different strength of the atom-dimer interaction. Note, however, that we neglect the atom-dimer scattering channel denoted by $\lambda_{3b}$ in Eq. \eqref{fullsu3trunc}. We again expect its effect to be subdominant.

\subsection{Flow equations and initial conditions}
The flow equations of the two-body sector are the straightforward generalization of Eqns. \eqref{mphifeq} and \eqref{aphifeq} with the replacements $m^2_\phi\to m^2_{\phi i}$, $A_\phi\to A_{\phi i}$, and $h\to h_i$. Since the Yukawa couplings $h_i$ are not renormalized the flow equations can be immediately integrated. As in section \ref{sec:5}, the microscopic values of the gap terms $m_{\phi i}^2$ and the $h_i$ are chosen to reproduce the experimentally measured scattering lengths $a_{12}, a_{23}, a_{13}$ in the infrared. Due to our ignorance of the microscopic details (e. g. effective range) leading to the specific scattering profiles in the regime between $B=100\, \textrm{G}$ and $B=550\, \textrm{G}$ (see Fig. \ref{fig:scattlength}), we choose here large and equal values of $h=h_1=h_2=h_3$. This corresponds to contact interactions at the UV scale. However, we use the dependence on the specific choice of $h$ to estimate the error within our truncation. We emphasize that in this problem the bosonic fields $\phi_i$ are not associated with the Feshbach molecules of the nearby resonance. They should rather be seen as true auxiliary Hubbard-Stratonovich fields representing the T-matrix of the atom-atom scattering.

The flow equations of the three-body sector read for the gap term,
\begin{equation}
\partial_t m_\chi^2 = \sum_{i=1}^3 \frac{2 g_i^2 k^5}{\pi^2 A_{\phi i}(3k^2-2\mu+2m_{\phi i}^2/A_{\phi i})^2}
\label{eq:flowofm}
\end{equation}
and for the Yukawa-like coupling $g_1$ one finds
\begin{equation}
\partial_t g_1  = -\frac{g_2 h_2 h_1 k^5\left(6k^2-5\mu+\frac{2m_{\phi2}^2}{A_{\phi 2}}\right)}{3\pi^2 A_{\phi 2}(k^2-\mu)^2\left(3k^2-2\mu+\frac{2m_{\phi2}^2}{A_{\phi2}}\right)^2}-\frac{g_3 h_3 h_1 k^5\left(6k^2-5\mu+\frac{2m_{\phi3}^2}{A_{\phi 3}}\right)}{3\pi^2 A_{\phi 3}(k^2-\mu)^2\left(3k^2-2\mu+\frac{2m_{\phi3}^2}{A_{\phi3}}\right)^2}.
\label{eq:flowofg}
\end{equation}
The flow equations for $g_2$ and $g_3$ can be derived from Eq. \eqref{eq:flowofg} by permuting the corresponding indices. There is an additional contribution to the flow of $g_i$ arising from the box diagram which regenerates the atom-dimer interaction during the RG flow. As explained in the previous section it can be computed using a $k$-dependent Hubbard-Stratonovich transformation and the flows of the $g_i$ receive the additional contribution
\begin{eqnarray}\label{referm}
m_\chi^2\left(-\frac{\partial_t \lambda_{ij}^{(3)}}{2 g_j}-\frac{\partial_t \lambda_{il}^{(3)}}{2 g_l}+\frac{g_i\partial_t \lambda_{jl}^{(3)}}{2 g_j g_l}\right).
\end{eqnarray}
Here we define $(i,j,l)=(1,2,3)$ and permutations thereof and find
\begin{eqnarray}
\partial_t\lambda_{ij}^{(3)}=\frac{k^5 h_1 h_2 h_3 h_l(9k^2-7\mu+\frac{4 m_{\phi l}^2}{A_{\phi l}})}{6\pi^2 A_{\phi l}(k^2-\mu)^3(3k^2-2\mu+\frac{2 m_{\phi l}^2}{A_{\phi l}})^2}.
\end{eqnarray}
Although this contribution is essential to obtain the full tower of Efimov states at the resonance it plays only a subdominant role for the properties of the lowest Efimov state. For simplicity we will therefore neglect this term in the computation of the loss coefficient $K_3$.

In the broad resonance limit, $h_i\to\infty$, Efimov physics is universal in the sense that it is only dependent on the given scattering lengths $a_{ij}$ and an additional three-body parameter \cite{BraatenHammer}. In section \ref{sec:5} we found that this parameter may be given by the cutoff scale $\Lambda$. Here, we however use the ratio of the couplings $g_i$ and $m_\chi^2$ at fixed UV scale $\Lambda$ as the three-body parameter, which is equivalent to using $\Lambda$ as the parameter itself. Also we choose $g=g_1=g_2=g_3$ at UV scale such that our three-body parameter is given by $\lambda^{(3)}\equiv-g^2(\Lambda)/m^2_\chi(\Lambda)$. Note, however, that the couplings $g_i$ evolve differently during the flow towards the IR.

\subsection{Numerical results and comparison with experiment}

We solve the flow equations \eqref{eq:flowofm} and \eqref{eq:flowofg} numerically. We note at this point that the zero-range limit is valid if the magnitude of the scattering length is much larger than the range of interaction given typically by the van der Waals length $l_{vdW}$, which for $^6\textrm{Li}$ is $l_{vdW}\approx 62.5\,a_0$. Although the loss resonances appear for $|a_{ij}|\gtrsim2l_{vdW}$ such that the zero-range approximation might be questionable, we will assume its validity. We will, however, estimate the error by using various values for the Yukawa coupling $h$, which is related to the effective range, as we have seen in section \ref{sec:5}.

The trimer energy level can be calculated in the same way as in the case of $\textrm{SU}(3)$ symmetric fermions by determining the chemical potential $\mu$ such that $m^2_\chi=0$ for $k=0$. We find $m^2_\chi=0$ for large enough values of $a_{ij}$ which indicates the presence of a trimer bound state. Now, the experimental observations offer a way to determine the three-body parameter inherent to Efimov physics: We simply adjust the initial value of $\lambda^{(3)}$ such that the Efimov trimer becomes degenerate with the atom threshold exactly at the position of the observed loss feature at $B=125\, \textrm{G}$. This fixes the three-body parameter $\lambda^{(3)}$ to the physical value and we assume that it is constant for all other investigated magnetic field strengths $B$.

\begin{figure}[t]
\centering
\includegraphics[width=0.5\textwidth]{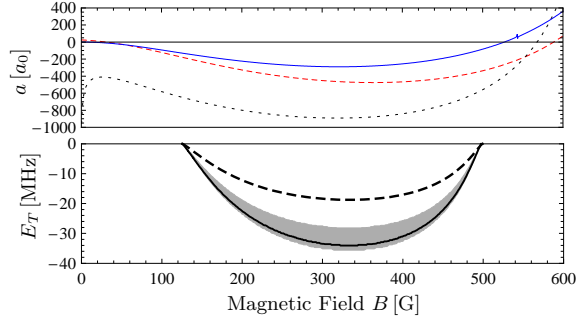}
\caption{{\itshape Upper panel:} Scattering length $a_{12}$ (solid), $a_{23}$ (dashed) and $a_{31}$ (dotted) as a function of the magnetic field $B$ for $^6$Li. {\itshape Lower panel:} Binding energy $E_T$ of the three-body bound state $\chi\widehat{=}\psi_1\psi_2\psi_3$. The solid line corresponds to the initial value $h^2=100\, a_0^{-1}$. The shaded region gives an error estimate by using values of $h$ in the range $h^2=20\, a_0^{-1}$ (upper border) to $h^2=300\, a_0^{-1}$ (lower border). The dashed line shows the reduction of the trimer energy when the refermionization of the atom-boson interaction is taken into account.}
\label{fig:Energies}
\end{figure}

Having determined the three-body parameter we are able to calculate the energy level of the trimer bound state as a first prediction. We find the existence of the trimer in the magnetic field range from $B=125\, \textrm{G}$ to $B=498\, \textrm{G}$. The binding energy $E_T$ is shown in Fig. \ref{fig:Energies} as solid line when choosing $h^2=100\, a_0^{-1}$ as appropriate for $^6\textrm{Li}$ in the $(1,2)$ channel. The shaded region is an error estimate obtained by choosing values $h^2\in (20\,a_0^{-1},300\, a_0^{-1})$. If we include the contribution due to the $k$-dependence of the Hubbard-Stratonovich transformation given in Eq. \eqref{referm} we find a reduction of the trimer binding energy. The result is shown as dashed curve.

\begin{figure}[t!]
\begin{minipage}[c]{0.3\textwidth}
\centering
\includegraphics[width=0.8\textwidth]{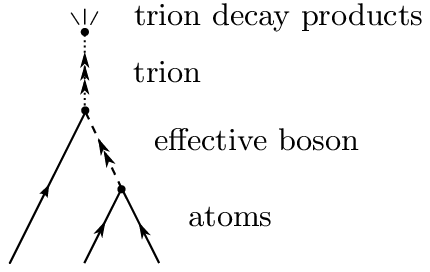}
\end{minipage}%
\begin{minipage}[c]{0.7\textwidth}
\centering 
\includegraphics[width=0.95\textwidth]{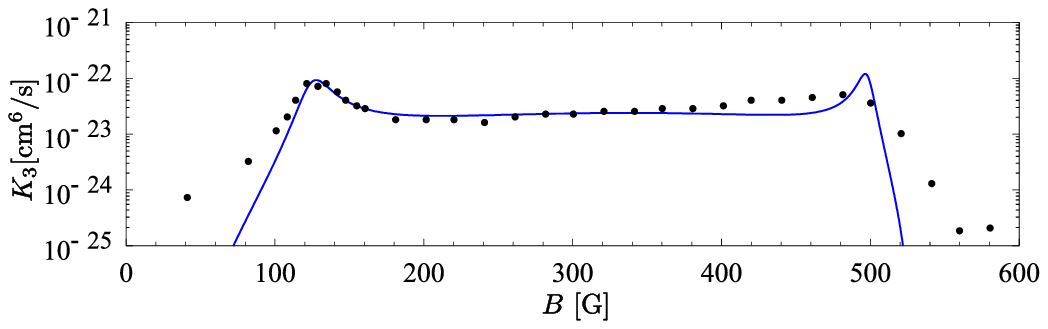}
\end{minipage}
\caption{\textit{left:} Three-body loss involving the Efimov trimer. \textit{right:} Loss coefficient $K_3$ in dependence on the magnetic field $B$ as measured in \cite{Ottenstein08} (dots). The solid line is the fit of our model to the experimental curve. We use here a decay width $\Gamma_\chi$ that is independent of the magnetic field $B$.}
\label{fig:K3}
\end{figure}

As a second prediction we determine the loss coefficient $K_3$ measured in experiments \cite{Ottenstein08,Huckans09}. 
Within our model the observed three-body loss can be explained as follows below, an illustration is shown in Fig. \ref{fig:K3} (left). So far we have shown the existence of the Efimov state in the regime of interest. This trimer state may the formed in a collision of three atoms. However, one must keep in mind that $^6$Li atoms have a complicated substructure with various bound states so that the composite trimer can become unstable. We account for this decay to not further specified degrees of freedom by a phenomenological decay width $\Gamma_\chi$. For example, the decay might involve the molecular dimer states associated with the nearby Feshbach resonance. The trimer formation from three atoms proceeds via the exchange of the effective bosons $\phi_i$. Evaluating the matrix elements corresponding to the diagram shown in Fig. \ref{fig:K3} (left) we can estimate the loss coefficient $K_3$ to be proportional to 
\begin{equation}
p=\left| \sum_{i=1}^3 \frac{h_i g_i}{m_{\phi i}^2}\frac{1}{\left(m_\chi^2-i \frac{\Gamma_\chi}{2}\right)}\right|^2.
\label{eq:decayprob}
\end{equation}
The amplitude for forming a trimer out of three atoms is given by $\sum_{i=1}^3 h_i g_i/m_{\phi i}^2$. The amplitude for the evolution from the initial state of three atoms to the unknown decay products further involves the inverse trimer propagator which in the real-time formalism reads
\begin{equation}
G_\chi^{-1}=\omega-\frac{\vec p^2}{3}-m_\chi^2+i\frac{\Gamma_\chi}{2}.
\end{equation}
We evaluate it in the limit of small momenta $\textbf p^2=(\sum_i \textbf p_i)^2\to0$, and small on-shell atom energies $\omega_i=\vec p_i^2$, $\omega=\sum_i \omega_i\to 0$. That the incoming atoms in the recombination process are at the atom threshold also implies that we have an effective chemical potential $\mu=0$. The trimer gap $m_\chi^2$ is then negative for $B=125\ldots498\, \textrm{G}$. A further ingredient of the calculation of the matrix elements in Eq. \eqref{eq:decayprob} is the decay width $\Gamma_\chi$ which we first assume to be independent of the magnetic field $B$. Furthermore the evaluation of $K_3$ via Fermi's Golden rule involves unknown vertices and phase-space factors for the trimer decay such that our compuation contains an unknown multiplicative factor $c_K$. In terms of $p$ in Eq. \eqref{eq:decayprob} we obtain
\begin{equation}
K_3=c_K \, p.
\end{equation}
The resulting curve is shown in Fig. \ref{fig:K3} (right) and agrees reasonably well with the experimentally measured data points. To sum up, our estimation of $K_3$ involved three parameters: The location of the first loss resonance at $B_0=125\, \textrm{G}$ fixes the three-body parameter. The overall amplitude of $K_3$ at the first resonance is adjusted by $c_K$ and the unkown decay width $\Gamma_\chi$ is tuned to the width of the first resonance yielding a trimer lifetime of $\approx10\, \textrm{ns}$. The extension of the curve away from the first resonance then involves no further parameter.

Our calculation predicts a second sharp resonance feature at $B=498\,\textrm G$ where the trimer becomes again degenerate with the atom threshold. The assumption of a constant $\Gamma_\chi$ fixes the width of this peak. The measured feature is however much broader. This may have a simple explanation: It is plausible to assume that the trimer decays into the molecular dimer states associated with the nearby Feshbach resonances and a free atom. Approaching the Feshbach resonances the molecular state becomes closer to the trimer energy level. Therefore it is reasonable to assume that the matrix element for the decay from the trimer to the dimers becomes larger. This in turn implies a shortening of the lifetime of the trimer as the magnetic field $B$ is increased. We have tested this assumption and indeed found a significant broadening or even disappearance of the second resonance peak. Recently, there had been phenomenological studies of this reasoning in \cite{Wenz09}.

By now, the system of three-component $^6\textrm{Li}$ atoms is quite well understood also away from the magnetic field range discussed above. This includes theoretical predictions \cite{Hammer10,Naidon10} as well as experimental measurements \cite{Williams09,Lompe10a} which also finally led to the first direct association of a Efimov trimer in ultracold atoms by Lompe \textit{et al.} \cite{Lompe10b}.

\section{Conclusions}
\label{sec:conclusions}
During recent years non-relativistic few-body physics has experienced a revived interest and undergone considerable progress. A wealth of experiments with ultracold atomic gases together with a significant theoretical effort initiated the (re)birth of an exciting field known as Efimov physics. In this paper we reviewed the studies of various aspects related to the Efimov effect using the method of functional renormalization.

By constructing solutions to exact functional renormalization group flow equations, we have obtained a rather detailed picture of non-relativistic few-body problems related to the Efimov effect. We considered identical bosons and fermions with two or three spin components, the latter both for the case of a common Feshbach resonance with a global $\textrm{SU}(3)$ symmetry and for the case of $^6$Li where this symmetry is broken.

Solving the flow equations in terms of a vertex expansion leads to (numerically) exact solutions. A remarkable hierarchy in the structure of the flow equations allows to construct the parts of the quantum effective action relevant for the one-, two- or three-body problem without further approximations. Beyond the present context it becomes apparent how the quantum mechanical formalism is embedded in non-perturbative quantum field theory.

Besides the vertex expansion we also established the derivative expansion as a way to find approximate solutions. In connection with a $k$-dependent Hubbard-Stratonovich transformation this allows for a simple and very efficient parameterization of the essential physics. No equivalent approximation scheme is known to us for the quantum mechanical formalism. As we have demonstrated this scheme is particularly valuable in more complex situations where a full solution in the vertex expansion becomes too involved. An example for this was given by the calculation of the three-body loss rate in a system of ultracold $^6\textrm{Li}$ atoms which shows how our theoretical investigations can be directly tested in experiments. 

Both in the vertex expansion and in the derivative expansion the Efimov effect shows up as a limit cycle in the solution of the flow equations at the point where the scattering length diverges. This mechanism is very interesting from a theoretical perspective. Realistic models in relativistic quantum field theory that show a similar behavior have not been found so far. Also the universality of the non-relativistic few-body physics is very interesting from a theoretical point of view. While in a renormalization group description universality is usually associated with the presence of fixed points, the Efimov effect with its limit cycle scaling and the connected discrete scaling symmetry goes beyond this standard paradigm. This generalized notion of few-body universality is expected to extend beyond the physics of three particles. First studies of four-body universality using functional renormalization have been undertaken \cite{RM}. 

In conclusion, we believe that the functional renormalization group is a powerful and suitable method for investigation of  non-relativistic few-body physics and of the Efimov effect in particular. It allows to address few-body problems from an unconventional perspective and provides an interesting alternative to the standard quantum mechanical and effective field theory treatments. Importantly, the method is not only limited to few-body problems but also is widely used for studying  many-body physics. In this way functional renormalization provides a unified framework for studying an interplay between few-particle and many-particle phenomena.

\begin{acknowledgements}
We thank M. Birse, E. Braaten, S. Diehl, H. Gies, H.-W. Hammer, H. C. Krahl, J. M. Pawlowski, M. M. Scherer, C. Wetterich, and W. Zwerger for useful discussion and collaboration. We also acknowledge S. Jochim and the members of his group for interesting discussions and providing us with their experimental data. The work of S.F. has been supported by the DFG research group FOR 723 and the Helmholtz Alliance HA216/EMMI. S.M. is thankful to the Klaus Tschira foundation and the DFG research group FOR 723 for support. R. S. thanks the DFG for support within the FOR 801 `Strong correlations in multiflavor ultracold quantum gases'.
\end{acknowledgements}


\begin{thebibliography}{} 
\bibitem{BraatenHammer}
E.~Braaten and H.~W.~Hammer: 
Universality in few-body systems with large scattering length, 
Phys. Rep. {\bf 428}, 259, (2006).

\bibitem{Efimov70}
V.~Efimov:
Energy levels arising from resonant two-body forces in a three-body system,
  Phys. Lett.  \textbf{33B}, 563 (1970);
V.~Efimov:
Energy levels of three resonantly-interacting particles,
  Nucl. Phys. A  \textbf{210}, 157 (1973).

\bibitem{Bedaque}
  P.~F.~Bedaque, H.~W.~Hammer and U.~van Kolck,
  Renormalization of the three-body system with short-range interactions,
  Phys.\ Rev.\ Lett.\  {\bf 82}, 463 (1999).


\bibitem{Ferlainorev}
F.~Ferlaino, R.~Grimm:
Forty years of Efimov physics: How a bizarre prediction turned into a hot topic,
Physics \textbf{3}, 9 (2010).


\bibitem{Ottenstein08}
T.~B.~Ottenstein, T.~Lompe, M.~Kohnen, A.~N.~Wenz, and S.~Jochim:
Collisional Stability of a Three-Component Degenerate Fermi Gas,
Phys.~Rev.~Lett. \textbf{101}, 203202 (2008).

\bibitem{Huckans09}
J.~H.~Huckans, J.~R.~Williams, E.~L.~Hazlett, R.~W.~Stites, and K.~M.~O'Hara:
Three-Body Recombination in a Three-State Fermi Gas with Widely Tunable Interactions,
Phys.~Rev.~Lett. \textbf{102}, 165302 (2009).


\bibitem{reviews}
J.~Berges, N.~Tetradis and C.~Wetterich:
Non-perturbative renormalization flow in quantum field theory and statistical physics,
Phys.\ Rept.\  {\bf 363} (2002) 223;
T.~R.~Morris:
Elements of the Continuous Renormalization Group,
Prog.\ Theor.\ Phys.\ Suppl.\  {\bf 131} (1998) 395;
K.~Aoki:
Introduction to the Non-Perturbative Renormalization Group and its Recent Applications,
Int.\ J.\ Mod.\ Phys.\ B {\bf 14} (2000) 1249;
C.~Bagnuls and C.~Bervillier:
Exact renormalization group equations: an introductory review,
Phys.\ Rept.\  {\bf 348} (2001) 91; 
J.~Polonyi: 
Lectures on the functional renormalization group method,
Central Eur.\ J.\ Phys.\  {\bf 1} (2003) 1; 
M.~Salmhofer and C.~Honerkamp:
Fermionic renormalization group flows - Technique and Theory,
Prog.\ Theor.\ Phys.\  {\bf 105} (2001) 1; 
B.~Delamotte:
An Introduction to the Nonperturbative Renormalization Group,
cond-mat/0702365;
B.~J.~Schaefer and J.~Wambach:
Renormalization Group Approach towards the QCD Phase Diagram,
Phys. Part. Nucl.{\bf 39} (2008), 1025;
J.~M.~Pawlowski:
Aspects of the Functional Renormalisation Group,
Annals Phys. {\bf 322} (2007), 2831;
O.~J.~Rosten:
Fundamentals of the Exact Renormalization Group,
arXiv:1003.1366.

\bibitem{Wetterich}
C.~Wetterich:
Exact evolution equation for the effective potential,
Phys.\ Lett.\ B {\bf 301} (1993) 90.

\bibitem{Diehl:2005ae}
S.~Diehl and C.~Wetterich:
Functional integral for ultracold fermionic atoms,
Nucl. Phys. B {\bf 770}, 206 (2007);  
Universality in phase transitions for ultracold fermionic atoms,
Phys.\ Rev.\ A  {\bf 73} (2006) 033615; 
S.~Diehl:
Universality in the BCS - BEC Crossover in Cold Fermion Gases, 
cond-mat/0701157.
 
\bibitem{DKS}
S.~Diehl, H.~C.~Krahl, M.~Scherer:
Three-body scattering from nonperturbative flow equations,
Phys.\ Rev.\ C\  {\bf 78}, 034001 (2008).

\bibitem{MFSW}
 S.~Moroz, S.~Floerchinger, R.~Schmidt and C.~Wetterich:
 Efimov effect from functional renormalization,
 Phys.\ Rev.\  A {\bf 79}, 042705 (2009).

\bibitem{DFGPW}
 S.~Diehl, S.~Floerchinger, H.~Gies, J.~M.~Pawlowski and C.~Wetterich: 
 Functional renormalization group approach to the BCS-BEC crossover,
 Ann. Phys. (Berlin) {\bf 522}, 615 (2010).

\bibitem{SFDoctoralThesis}
S. Floerchinger, {\itshape Functional renormalization and ultracold quantum gases}, Doctoral thesis, Universit\"at Heidelberg, 2009.

\bibitem{Bosonization}
H.~Gies, C.~Wetterich, 
Renormalization flow of bound states,
Phys.~Rev.~D {\bf 65}, 065001 (2002); 
Renormalization flow from UV to IR degrees of freedom,
Acta~Phys.~Slov. {\bf 52}, 215 (2002).


\bibitem{FWCompositeOperators}
S. Floerchinger and C. Wetterich:
Exact flow equation for composite operators,
Phys. Lett. B {\bf 680}, 371 (2009).

\bibitem{SFBoundStates}
S. Floerchinger:
Exact flow equation for bound states,
Eur. Phys. J. C {\bf 69}, 119 (2010).

\bibitem{Birse08}
M.~C.~Birse:
Functional renormalization group for two-body scattering,
Phys. Rev. C {\bf 77}, 047001 (2008).

\bibitem{STM}
G.~V.~Skorniakov, K.~A.~Ter-Martirosian:
Three body problem for short range forces 1. Scattering of low energy neutrons by deuterons,
  Zh.~Eksp.~Teor.~Phys. \textbf{31}, 775 (1956), [Sov.~Phys.~JETP \textbf{4}, 648 (1957)].


\bibitem{Zwerger}
I.~Bloch, J.~Dalibard, and W.~Zwerger:
Many-body physics with ultracold gases,
Rev.~Mod.~Phys. {\bf 80}, 885 (2008).

\bibitem{Pitaevskii}
S.~Giorgini, L.~P.~Pitaevskii, and S.~Stringari:
Theory of ultracold atomic Fermi gases,
Rev.~Mod.~Phys. {\bf 80}, 1215 (2008).
 
 

\bibitem{BCSBEC}
M.~C.~Birse, B.~Krippa, J.~A.~McGovern, and N.~R.~Walet:
Pairing in many-fermion systems: an exact renormalisation group treatment,
Phys. Lett. B {\bf 605}, 287 (2005);
S.~Diehl, H.~Gies, J.~M.~Pawlowski, and C.~Wetterich: 
Flow equations for the BCS-BEC crossover,
Phys. Rev. A {\bf 76}, 021602(R) (2007); 
Renormalization flow and universality for ultracold fermionic atoms,
Phys. Rev. A 76, 053627 (2007); 
S.~Floerchinger, M.~Scherer, S.~Diehl, and C.~Wetterich:
Particle-hole fluctuations in BCS-BEC crossover,
Phys. Rev. B {\bf 78}, 174528 (2008);
L.~Bartosch, P.~Kopietz, and A.~Ferraz:
Renormalization of the BCS-BEC crossover by order-parameter fluctuations,
Phys. Rev. B {\bf 80}, 104514 (2009).

\bibitem{FSW}
S.~Floerchinger, M.~M.~Scherer, and C.~Wetterich:
Modified Fermi sphere, pairing gap, and critical temperature for the BCS-BEC crossover,
Phys. Rev. A {\bf 81}, 063619 (2010).

\bibitem{Petrov}
D.~S.~Petrov, C.~Salomon, and G.~V.~Shlyapnikov:
Weakly Bound Dimers of Fermionic Atoms,
Phys. Rev. Lett. {\bf 93}, 090404 (2004).

\bibitem{Manchester1}
B.~Krippa, N.~R.~Walet, M.~C.~Birse:
Renormalization group, dimer-dimer scattering, and three-body forces,
Phys. Rev. A {\bf 81}, 043628 (2010).

\bibitem{Manchester2}
M.~C.~Birse, B.~Krippa, N.~R.~Walet:
Convergence of a renormalization group approach to dimer-dimer scattering, 
arXiv:1011.5852.

\bibitem{SFW}
M.~M.~Scherer, S.~Floerchinger, and H.~Gies:
Functional renormalization for the BCS-BEC crossover,
e-print arXiv:1010.2890.


\bibitem{FSMW}
 S.~Floerchinger, R.~Schmidt, S.~Moroz, and C.~Wetterich:
 Functional renormalization for trion formation in ultracold fermion gases,
Phys.\ Rev.\  A {\bf 79}, 013603 (2009).
 
 \bibitem{Kokkelmans}
S.~J.~J.~M.~F.~Kokkelmans, J.~N.~Milstein, M.~L.~Chiofalo, R.~Walser, and M.~J.~Holland:
Resonance superfluidity: Renormalization of resonance scattering theory,
Phys.~Rev.~A {\bf 65}, 053617  (2002).



\bibitem{Bartenstein}
M.~Bartenstein, A.~Altmeyer, S.~Riedl, R.~Geursen, S.~Jochim, C.~Chin, J.~Hecker Denschlag, R.~Grimm,
A.~Simoni, E.~Tiesinga, C.~J.~Williams, and P.~S.~Julienne:
Precise Determination of 6Li Cold Collision Parameters by Radio-Frequency Spectroscopy on Weakly Bound Molecules,
Phys.~Rev.~Lett. {\bf 94}, 103201 (2005).

\bibitem{RM}
R.~Schmidt, and S.~Moroz:
Renormalization-group study of the four-body problem,
Phys.~Rev.~A {\bf 81}, 052709 (2010).


\bibitem{Honerkamp}
A.~Rapp, G.~Zarand, C.~Honerkamp, and W.~Hofstetter:
Color Superfluidity and ``Baryon'' Formation in Ultracold Fermions,
Phys.~Rev.~Lett. {\bf 98}, 160405 (2007);
A.~Rapp, W.~Hofstetter, and G.~Zarand:
Trionic phase of ultracold fermions in an optical lattice: A variational study,
Phys.\ Rev.\ B {\bf 77}, 144520 (2008).

\bibitem{Wilczek}
F. Wilczek, 
Quantum chromodynamics: Lifestyles of the small and simple,
Nature Phys. {\bf 3}, 375 (2007).


\bibitem{FScW}
 S.~Floerchinger, R.~Schmidt, and C.~Wetterich:
 Three-body loss in lithium from functional renormalization,
 Phys.\ Rev.\ A {\bf 79}, 053633 (2009).

\bibitem{Braatenli6}
  E.~Braaten, H.~W.~Hammer, D.~Kang and L.~Platter:
  Three-Body Recombination of $^6$Li Atoms with Large Negative Scattering Lengths,
Phys.~Rev.~Lett. \textbf{103}, 073202 (2009).

\bibitem{Naidon}
P.~Naidon and M.~Ueda:
Possible Efimov Trimer State in a Three-Hyperfine-Component Lithium-6 Mixture,
Phys.~Rev.~Lett. \textbf{103}, 073203 (2009).

\bibitem{Wenz09}
A.~N.~Wenz, T.~Lompe, T.~B.~Ottenstein, F.~Serwane, G.~Z\"urn, and S.~Jochim:
Universal trimer in a three-component Fermi gas,
Phys.~Rev.~A {\bf 80}, 040702 (R) (2009).

\bibitem{Hammer10}
H.~W.~Hammer, D.~Kang and L.~Platter:
Efimov physics in atom-dimer scattering of $^6$Li atoms,
Phys.~Rev.~A \textbf{82}, 022715 (2010).

\bibitem{Naidon10}
P.~Naidon and M.~Ueda:
The Efimov effect in lithium 6, 
\textit{eprint:} arXiv:1008.2260v2 (2010).


\bibitem{Williams09}
J.~R.~Williams, E.~L.~Hazlett, J.~H.~Huckans, R.~W.~Stites, Y.~Zhang, and K.~M.~O'Hara:
Evidence for an Excited-State Efimov Trimer in a Three-Component Fermi Gas,
Phys.~Rev.~Lett. \textbf{103}, 130404 (2009).

\bibitem{Lompe10a}
T.~Lompe, T.~B.~Ottenstein, F.~Serwane, K.~Viering, A.~N.~Wenz, G.~Z\"urn, and S.~Jochim:
Atom-Dimer Scattering in a Three-Component Fermi Gas,
Phys.~Rev.~Lett. \textbf{105}, 103201 (2010).


\bibitem{Lompe10b}
T.~Lompe, T.~B.~Ottenstein, F.~Serwane, A.~N.~Wenz, G.~Z\"urn, and S.~Jochim:
Radio-Frequency Association of Efimov Trimers,
Science {\bf 330}, 940 (2010).

\end{thebibliography}
\end{document}